\documentclass[acmsmall,screen,nonacm]{acmart}

\usepackage{tcolorbox}
\newcommand{\rqbox}[1]{\begin{tcolorbox}[colback=gray!10,colframe=black!40,boxrule=0.5pt,left=1mm,right=1mm,top=0.5mm,bottom=0.5mm]
#1
\end{tcolorbox}}

\usepackage{booktabs}
\usepackage{tabularx}
\usepackage{listings}
\usepackage{xcolor}

\usepackage{colortbl}
\usepackage{graphicx}
\usepackage{tikz}
\usepackage{amsthm}
\usepackage[capitalise]{cleveref}

\theoremstyle{acmdefinition}

\usetikzlibrary{arrows.meta,positioning,calc,fit,backgrounds,shapes.geometric}
\pgfdeclarelayer{background}
\pgfsetlayers{background,main}

\usepackage[htt]{hyphenat}
\usepackage{xspace}

\usepackage[normalem]{ulem}
\usepackage{framed}
\usepackage[inline]{enumitem}
\setitemize{noitemsep,topsep=1pt,parsep=1pt,partopsep=1pt}

\newcommand{\numCertAutoNoSamplePct}{91.3}%

\newcommand{\numChosenCertsPct}{93.7}%

\newcommand{\numCorpusChangedPct}{16.3}%

\newcommand{\numGtSynthRecall}{98.0}%

\newcommand{\numGtVbRecall}{91.6}%

\newcommand{\numStrongAutoPct}{87.7}%

\newcommand{\numSynthChangedPct}{9.2}%

\newcommand{\numToolCpuHoursPerProcess}{4.8}%

\newcommand{\numVbSingleWeakChangedPct}{7.5}%

\newcommand{\numVbSingleStrongChangedPct}{70.2}%

\newcommand{\numVeribenchChangedPct}{16.7}%

\newcommand{\numCorpusFiles}{2{,}608}%

\newcommand{\numWeakAutoPct}{91.9}%

\newcommand{\numAgentPromptExamplesTotal}{7}%
\newcommand{\numAgentPromptExamples}{4}%
\newcommand{\numAgentClosed}{336}%
\newcommand{\numAgentCostUsd}{90.09}%
\newcommand{\numAgentCostPerMethodUsd}{0.27}%
\newcommand{\numAgentWallHours}{6.8}%
\newcommand{\numAgentMedianTurns}{10}%
\newcommand{\numAgentFirstSubmissionPass}{279}%
\newcommand{\numAgentMaxValidatorCalls}{10}%
\newcommand{\numAgentFullCorpusUsd}{1{,}048}%
\newcommand{\numAgentFullCorpusHours}{79}%

\newcommand{\numUncertified}{340}%

\newcommand{\numCertAutoNoSample}{3{,}570}%

\newcommand{\numVeribenchTotal}{3{,}692}%
\newcommand{\numVeribenchChanged}{618}%
\newcommand{\numSynthTotal}{218}%
\newcommand{\numSynthChanged}{20}%
\newcommand{\numSampleTotal}{491}%
\newcommand{\numSampleWeak}{431}%
\newcommand{\numCorpusTotal}{3{,}910}%
\newcommand{\numCorpusStrong}{587}%
\newcommand{\numCorpusWeak}{3{,}323}%
\newcommand{\numCorpusChanged}{638}%

\newcommand{\numStrongAuto}{515}%

\newcommand{\numWeakAutoOf}{3{,}323}%

\newcommand{\numWeakAuto}{3{,}055}%

\newcommand{\numSynthToWeak}{19}%
\newcommand{\numSynthToStrong}{1}%

\newcommand{\numWeakToStrongReviewed}{10}%
\newcommand{\numWeakToStrongLabelWrong}{10}%
\newcommand{\numWeakToStrongAgreed}{10}%

\newcommand{\numVbSingleToWeak}{252}%
\newcommand{\numVbSingleToStrong}{88}%
\newcommand{\numVbSingleStrongTotal}{359}%
\newcommand{\numVbSingleWeakTotal}{1{,}175}%
\newcommand{\numVbMultiMethods}{2{,}158}%

\newcommand{\numVbSingleFiles}{1{,}534}%

\newcommand{\numExcludedFileVerify}{2{,}612}%
\newcommand{\numExcludedTotal}{2{,}880}%

\newcommand{\numFigTimeStrongMethods}{515}%
\newcommand{\numFigTimeStrongHours}{0.3}%
\newcommand{\numFigTimeWeakMethods}{3{,}055}%
\newcommand{\numFigTimeWeakHours}{42.1}%
\newcommand{\numFigTimeUncertifiedMethods}{340}%
\newcommand{\numFigTimeStrongMedianSeconds}{2.6}%
\newcommand{\numFigTimeWeakMedianSeconds}{35.0}%
\newcommand{\numFigTimeUncertifiedHours}{15.6}%
\newcommand{\numFigTimeUncertifiedSharePct}{26.8}%
\newcommand{\numFigScoreWeakFull}{21}%
\newcommand{\numFigScoreStrongBelow}{61}%

\newcommand{\numPrecondConjunctsPct}{88.3}%

\newcommand{\numPrecondPartPct}{99.0}%

\newcommand{\numPrecondFullPct}{74.9}%

\newcommand{\numChosenFiles}{404}%
\newcommand{\numChosenCoverage}{100.0}%
\newcommand{\numChosenPct}{22.7}%
\newcommand{\numChosenOps}{12}%

\newcommand{\numGtVbUniverse}{2{,}841}%

\newcommand{\numAblationMaxExamples}{200}%
\newcommand{\numAblationGuided}{3{,}771}%
\newcommand{\numAblationUnguided}{3{,}432}%
\newcommand{\numAblationGuidedPct}{96.4}%
\newcommand{\numAblationUnguidedPct}{87.8}%
\newcommand{\numAblationGuidedOnly}{346}%
\newcommand{\numAblationGuidedFull}{3{,}512}%
\newcommand{\numAblationUnguidedFull}{2{,}942}%
\newcommand{\numAblationGuidedAcceptPct}{100.0}%
\newcommand{\numAblationUnguidedAcceptPct}{11.8}%

\newcommand{\numResidueWeak}{268}%
\newcommand{\numResidueStrong}{72}%
\newcommand{\numResidueWeakNoInput}{107}%
\newcommand{\numResidueWeakNoDivergence}{145}%
\newcommand{\numResidueWeakGeneratorError}{9}%
\newcommand{\numResidueWeakNoMutants}{7}%
\newcommand{\numResidueWeakAmpleButSilent}{96}%
\newcommand{\numResidueWeakStarved}{49}%
\newcommand{\numBudgetMutationSeconds}{300}%
\newcommand{\numBudgetResourceMillions}{250}%
\newcommand{\numResidueStrongProofLines}{32}%
\newcommand{\numResidueStrongProofLinesMax}{80}%
\newcommand{\numResidueStrongLemma}{59}%
\newcommand{\numResidueStrongAssert}{69}%
\newcommand{\numResidueStrongLoopAnnotation}{34}%
\newcommand{\numResidueStrongProofSeconds}{3.7}%

\newcommand{\numToolTimeout}{97}%
\newcommand{\numPrecondMethods}{3{,}031}%
\newcommand{\numPrecondConjuncts}{10{,}242}%
\newcommand{\numPrecondNoneMethods}{30}%

\newcommand{\numToolConcurrency}{12}%

\newcommand{\numCleanSubset}{3{,}570}%
\newcommand{\numBaselineMccAll}{0.6569}%
\newcommand{\numBaselineMccClean}{0.8837}%
\newcommand{\numBaselineCpuHours}{2.1}%
\newcommand{\numBaselineSureStrong}{453}%
\newcommand{\numProofPulseMccAll}{0.4881}%
\newcommand{\numProofPulseMccClean}{0.5228}%
\newcommand{\numProofPulseCpuHours}{4.2}%
\newcommand{\numReducedMccAll}{0.9004}%
\newcommand{\numReducedMccClean}{0.9226}%
\newcommand{\numReducedError}{16}%
\newcommand{\numReducedCpuHours}{58.0}%
\newcommand{\numReducedHeur}{324}%

\newcommand{\numScoreScored}{324}%
\newcommand{\numScoreStrong}{72}%
\newcommand{\numScoreAuc}{0.8725}%
\newcommand{\numScoreBestThreshold}{83}%
\newcommand{\numScoreBestMcc}{0.6955}%
\newcommand{\numScoreThresholdMcc}{0.0968}%

\newcommand{\numAbFiles}{491}%
\newcommand{\numAbUnrestrictedHours}{20.7}%
\newcommand{\numAbUnrestrictedMutationHours}{16.0}%
\newcommand{\numAbReducedHours}{9.5}%
\newcommand{\numAbReducedMutationHours}{6.4}%
\newcommand{\numAbSpeedup}{2.2}%
\newcommand{\numAbUnrestrictedTimeout}{96}%
\newcommand{\numAbReducedTimeout}{17}%
\newcommand{\numAbUnrestrictedWitnesses}{381}%
\newcommand{\numAbReducedWitnesses}{402}%
\newcommand{\numAbBothWitnesses}{377}%
\newcommand{\numAbGained}{25}%
\newcommand{\numAbLost}{4}%
\newcommand{\numAbLostToTimeout}{2}%

\newcommand{\numHeldOutFiles}{50}%
\newcommand{\numPublishedSampleCovered}{394}%
\newcommand{\numPublishedSamplePct}{97.5}%
\newcommand{\numAddedOnlySample}{10}%
\newcommand{\numAddedOnlySamplePct}{2.5}%
\newcommand{\numAddedOnlyHeldOut}{8}%
\newcommand{\numAddedOnlyHeldOutPct}{16.0}%
\newcommand{\numPublishedFloor}{97.5}%
\newcommand{\numPublishedCoverOps}{8}%
\newcommand{\numPublishedCoverPct}{52.5}%
\newcommand{\numMatchedCoverOps}{7}%
\newcommand{\numMatchedCoverPct}{8.0}%
\newcommand{\numPublishedCostRatio}{6.52}%

\newcommand{\numWeakToStrongSameAsStrongSpec}{4}%
\newcommand{\numWeakToStrongStrongSpecNoEnsures}{5}%

\newcommand{\numToolCpuHours}{58.0}%

\newcommand{\numVeribenchToWeak}{315}%
\newcommand{\numVeribenchToStrong}{303}%

\newcommand{\numVbSingleChanged}{340}%

\newcommand{\numResidueWeakStarvedNoSearch}{16}%

\definecolor{cProof}{HTML}{1C7A58}
\definecolor{cWit}{HTML}{A8511A}
\definecolor{cAgent}{HTML}{2F4B7C}
\definecolor{cValid}{HTML}{555555}
\definecolor{cMuted}{HTML}{69748A}
\definecolor{cRule}{HTML}{C9CFDA}

\newcommand{\circnum}[1]{\textcolor{cAgent}{\textbf{\textcircled{\scriptsize #1}}}}

\lstdefinelanguage{Dafny}{
  morekeywords={method,function,predicate,lemma,returns,requires,ensures,
    modifies,reads,decreases,invariant,assert,assume,expect,var,if,else,while,for,
    forall,exists,in,new,old,multiset,seq,set,map,array,array2,int,bool,nat,
    char,string,module,import,opened,return,print,ghost,datatype,type,true,false},
  sensitive=true,
  morecomment=[l]{//},
  morecomment=[s]{/*}{*/},
  morestring=[b]",
  morestring=[b]',
}
\lstdefinestyle{certtarget}{backgroundcolor=\color{cRule!35}, rulecolor=\color{cMuted}}
\lstdefinestyle{certpart}{aboveskip=1pt}
\newcommand{\certtag}[2]{\par\noindent{\fontsize{5.6}{6.4}\selectfont\sffamily\textcolor{#1}{#2}}\par}

\newcommand{\tool}{\textsf{DafnyCert}\xspace}

\newcommand{\code}[1]{\texttt{\small #1}}

\newcommand{\datasetBig}{\textsf{VeriEquivBench}\xspace}
\newcommand{\veqcheck}{\textsf{VEB-Check}\xspace}

\newcommand{\datasetSmall}{\textsf{Dafny-Synthesis}\xspace}

\begin{document}

\title[\tool]{\tool: Certifying Specification Strength in Dafny}

\author{Álvaro Silva}
\orcid{0009-0005-2941-9942}
\affiliation{%
  \institution{Faculdade de Engenharia, Universidade do Porto}
  \city{Porto}
  \country{Portugal}
}
\email{amfpsilva@hotmail.com}

\author{Isabel Amaral}
\orcid{0009-0009-4188-9233}
\affiliation{%
  \institution{Faculdade de Engenharia, Universidade do Porto}
  \city{Porto}
  \country{Portugal}
}
\email{isabel.andre.amaral@gmail.com}

\author{Ruben Martins}
\orcid{0000-0003-1525-1382}
\affiliation{%
  \institution{Carnegie Mellon University}
  \department{Computer Science Department}
  \city{Pittsburgh}
  \country{USA}
}
\email{rubenm@andrew.cmu.edu}

\author{Alexandra Mendes}
\orcid{0000-0001-8060-5920}
\affiliation{%
  \institution{Faculdade de Engenharia, Universidade do Porto}
  \city{Porto}
  \country{Portugal}
}
\email{alexandra@archimendes.com}

\begin{abstract}

Formally verifying a method with Dafny guarantees correctness relative to its specification.
However, a specification may be too permissive, allowing multiple outputs for the same input, even when the implementation produces only one. In this scenario, a different implementation could produce a different result while still satisfying the same specification. Therefore,
successful verification does not necessarily establish that the method behaves as intended. 

In this paper, we propose checkable certificates that provide independently verifiable evidence of whether a specification is weak or strong. A \emph{certificate of weakness} contains an alternative implementation that verifies against the original specification, together with a witness input on which the original and alternative implementations produce different results. A \emph{certificate of strength} is produced by constructing a harness around the original method such that successfully verifying the harness guarantees that the specification allows at most one possible output for any given input.

We present \tool, a tool that automatically produces these certificates. Strength is established by constructing and verifying the harness. Weakness is demonstrated by using the MutDafny mutation tool to find mutants that verify against the original specification, and by using the Hypothesis property-based testing framework to generate inputs that produce different outputs for the original program and the mutant. The resulting mutant and witness form a recheckable certificate of weakness. 
We reduce the cost of mutation-based search by selecting a smaller set of operators that generates only \numChosenPct{}\% of the mutants on the selection set, covering \numGtVbRecall{}\% of the \numGtVbUniverse{} weak specifications of \datasetBig{} in an independent transfer set.

On a dataset of \numCorpusTotal{} verified Dafny methods from \datasetSmall{} and \datasetBig{}, \tool{} automatically certifies \numCertAutoNoSample{} methods (\numCertAutoNoSamplePct{}\%). We also provide certificates for the remaining 340 methods, all accepted by the same validator. To our knowledge, this is the first dataset in which every specification is backed by a checkable certificate of strength or weakness.
 
These labels disagree with the benchmarks' original labels for \numCorpusChangedPct{}\% of the methods. Among single-method \datasetBig{} files, \numVbSingleStrongChangedPct{}\% of the specifications labeled strong are certified weak, and in \datasetSmall{}, whose labels were manually assigned, \numSynthToWeak{} of the \numSynthChanged{} disagreements also correspond to specifications labeled strong but certified weak.
These findings highlight the importance of supporting specification strength labels, whether assigned by humans or automated methods, with independently checkable evidence.

\end{abstract}




\maketitle

\section{Introduction}
\label{sec:intro}

Formal verification has been used to establish the functional correctness of real systems, such as an operating system kernel \cite{sel4}, a C compiler \cite{ccompiler}, distributed systems \cite{ironfleetdistributed}, and a cryptographic library \cite{cryptLib}. Verification-aware languages such as Dafny \cite{dafnyLeino}, Why3 \cite{why3}, and Verus \cite{verus} bring these techniques to everyday code, where the developer provides \emph{preconditions (requires)} and \emph{postconditions (ensures)} for a method, and an SMT solver \cite{z3_paper} checks that the implementation satisfies the specification.

Successful verification establishes that an implementation satisfies its given specification, but does not ensure that the specification accurately captures the intended behavior. When a specification is too permissive, it can be satisfied by multiple implementations with different behaviors. Thus, the verifier may accept implementations whose behavior differs from the programmer's intent.

This problem arises in real verified systems. Fonseca et
al.~\cite{fonsecaDistributed} studied three formally verified distributed
systems and identified faults in their specifications.
In IronFleet \cite{ironfleetdistributed}, a distributed system written in Dafny, 
the specification of the replicated state machine did not
require exactly-once semantics, even though the implementation provided them. Consequently, the authors managed to disable request deduplication with a small patch that inserted a bug, but still verified against the original specification.
The authors conclude that formal verification requires techniques to test specifications.
Specification bugs have since been reported in other verified systems. CompCert's PowerPC semantics did not constrain the width of an immediate field~\cite{ccompBugs},
and IronSpec~\cite{ironspec} uncovered ten specification bugs in six verified Dafny systems, each of which admitted an incorrect implementation that still verified.

This problem has become increasingly relevant as LLMs are used to generate both code and specifications in verification-aware languages. Existing benchmarks for this task, such as \datasetSmall{}~\cite{dafny_synthesys_FSE24} and \datasetBig{}~\cite{VeriEquivBench}, classify specifications as strong or weak, providing labels for evaluating specification generation and assessment tools.

However, these labels are assigned by human annotators or LLMs and may contain errors. Prior work on \datasetBig{}~\cite{VeriEquivBench} introduced a strong specification checker, which we refer to as \veqcheck{}, that can automatically prove that a specification is strong, but cannot establish that a specification is weak. Other approaches, including mutation scores~\cite{Mutdafny,le-cong2025}, proof coverage~\cite{ProofPulse_paper}, and LLM judges~\cite{Clover}, provide only heuristic assessments rather than checkable guarantees.

We argue that specification strength should be determined by independently verifiable evidence. A specification is \emph{weak} if two implementations that verify against it return different outputs for the same valid input. In this work, to show weakness, we provide a second implementation and an input on which the output differs from that of the original implementation. Running both implementations on that input provides concrete evidence that the specification is weak. A specification is \emph{strong} when no such pair exists. To show strength, we provide a harness that is verifiable by Dafny only if the specification allows at most one output for each input. We call each of these self-contained artifacts a \emph{certificate}. Both certificate types, \emph{weak} and \emph{strong}, can be checked with \texttt{dafny verify} and \texttt{dafny run}, without trusting any of the tools used to generate them.

We built \tool{} to automatically generate these certificates. It first attempts to prove that a specification is strong. If it fails, it searches for evidence of weakness by mutating the implementation, retaining the verified mutants, and running them on inputs generated from the precondition.

\vskip 0.5em
\noindent
The main \textbf{contributions} of this paper are:
\begin{itemize}[leftmargin=*]
\item[\small{$\blacktriangleright$}] \textbf{A fully certified benchmark dataset for evaluating specification strength in Dafny.}
We provide certificates of strength or weakness for \numCorpusTotal{} methods from the \datasetSmall \cite{dafny_synthesys_FSE24} and \datasetBig \cite{VeriEquivBench} datasets. Each carries either an executable counterexample showing that the specification is weak (\emph{certified weak}), or a verifiable proof that it is strong (\emph{certified strong}). The certified labels differ from those in the existing benchmarks for \numSynthChangedPct{}\% (\numSynthChanged{}/\numSynthTotal{}) of \datasetSmall and \numVeribenchChangedPct{}\% (\numVeribenchChanged{}/\numVeribenchTotal{}) of \datasetBig.

\item[\small{$\blacktriangleright$}] \textbf{An automated method to certify specifications as weak.} We search for an alternative implementation by mutating the method, verifying each mutant against the \emph{unchanged} contract, and running the mutants that verify on inputs generated from the method's preconditions. 
 
This is enough to automatically certify \numWeakAutoPct{}\% (\numWeakAuto{}/\numWeakAutoOf{}) of the weak specifications.

\item[\small{$\blacktriangleright$}] \textbf{A certificate template and validator.} 
The validator checks certificate integrity and 

re-checks the evidence using Dafny. Trusting a certificate generated from a template only requires trusting the validator and Dafny.

\item[\small{$\blacktriangleright$}] \textbf{\tool, an end-to-end framework for sound specification strength certification.} By extending \veqcheck{} 

and integrating our approach for certifying specification weakness, \tool generates certificates for \numCertAutoNoSamplePct{}\% (\numCertAutoNoSample{}/\numCorpusTotal{}) of all methods across \datasetSmall and \datasetBig. The remaining certificates that complete the dataset, pass the same validator, making the final benchmark fully labeled by re-checkable evidence. 

\end{itemize}

\section{DafnyCert}
\label{sec:dafny_cert}

\begin{figure}[t]
\centering
\resizebox{\linewidth}{!}{%
\hyphenpenalty=10000\exhyphenpenalty=10000%
\begin{tikzpicture}[
  font=\Large,
  stepbox/.style={rectangle, rounded corners=2pt, draw=cRule, very thick,
    align=center, text width=2.05cm, minimum height=1.15cm, inner sep=3pt,
    fill=white},
  weakbox/.style={stepbox, draw=cWit, fill=cWit!7},
  spoolbox/.style={stepbox, draw=cWit, dashed, fill=cWit!18},
  strongbox/.style={stepbox, draw=cProof, fill=cProof!7},
  ctrlbox/.style={stepbox, draw=cAgent, fill=cAgent!12},
  valbox/.style={stepbox, draw=cValid, fill=cValid!18},
  iobox/.style={rectangle, rounded corners=2pt, draw=black, thick,
    align=center, text width=2.3cm, minimum height=1.5cm, fill=cRule!25,
    font=\bfseries\Large},
  decision/.style={diamond, aspect=1, draw=cValid, very thick, fill=white,
    inner sep=0pt, minimum width=9.5mm, minimum height=9.5mm,
    font=\Large\bfseries, text=cValid},
  flow/.style={-{Latex[length=2.2mm]}, thick},
  elabel/.style={font=\normalsize, fill=white, inner sep=1.5pt},
  ylab/.style={font=\normalsize\bfseries, text=cProof, inner sep=1pt,
    fill=white},
  nlab/.style={font=\normalsize\bfseries, text=cWit, inner sep=1pt,
    fill=white},
  hop/.style={flow, preaction={draw=white, line width=3.6pt}},
  lanelabel/.style={font=\Large\bfseries\itshape, fill=white, inner sep=2pt},
  lanebox/.style={rounded corners=3pt, dashed, thick, inner sep=3.4mm},
]

\node[weakbox, anchor=west, text width=3.3cm] (w1) at (0,0)
  {\circnum{3}~Hypothesis: sample admissible inputs, record $M$'s outputs};
\node[weakbox, right=3mm of w1, text width=3cm] (w2)
  {\circnum{4}~MutDafny: mutate $M$, verify each mutant against $S$};
\node[spoolbox, right=3mm of w2, text width=2.1cm] (w3)
  {\circnum{5}~alive mutants};
\node[weakbox, right=3mm of w3, text width=3cm] (w4)
  {\circnum{6}~run alive mutants on the sample, compare outputs};
\node[weakbox, right=3mm of w4, text width=2.9cm] (w5)
  {\circnum{7}~build a certificate per divergence, check it};
\node[decision, right=5mm of w5] (dw) {C};
\node[weakbox, right=10mm of dw, text width=3.4cm] (w6)
  {\circnum{8}~mutation score heuristic classification weak / 
   strong};

\draw[flow] (w1) -- (w2);
\draw[flow] (w2) -- (w3);
\draw[flow] (w3) -- (w4);
\draw[flow] (w4) -- (w5);
\draw[flow] (w5) -- (dw);
\draw[flow] (dw) -- node[above, nlab] {no} (w6);

\coordinate (cby) at ($(w5.south)+(0,-0.6)$);
\coordinate (cbk) at (dw.south|-cby);
\draw[flow, dashed] (dw.south) -- (cbk) -| (w2.south);
\node[elabel] (note1) at (9.6,0|-cbk)
  {on \emph{yes}, stop \circnum{4} and \circnum{6}};
\node[font=\normalsize\itshape, text=cMuted, fill=white, inner sep=1.5pt,
      below=1.2mm of note1] (note2)
  {\circnum{4} and \circnum{6} run concurrently: \circnum{6} reads
   \circnum{5} while \circnum{4} fills it};

\def\topY{4.1}
\node[iobox, anchor=west] (input) at (0,\topY)
  {method $M$\\+ spec $S$};

\node[strongbox, anchor=west, text width=2.6cm] (s1) at (3.2,\topY)
  {\circnum{1}~generate harness $H(S)$};
\node[strongbox, right=4mm of s1, text width=2.9cm] (s2)
  {\circnum{2}~verify $H(S)$ with Dafny};
\node[decision, right=4mm of s2] (ds) {C};
\draw[flow] (s1) -- (s2);
\draw[flow] (s2) -- (ds);

\node[iobox, text width=3.0cm] (unout) at (w6.center|-input.center)
  {uncertified\\{\normalsize\mdseries heuristic label only}};
\node[valbox, right=9mm of ds, text width=3.2cm] (val)
  {\circnum{9}~Validator re-checks with Dafny, rejects tampering};
\node[iobox, right=6mm of val] (out)
  {certified weak\\/ strong};

\draw[flow] (input) -- (s1);

\draw[flow] (ds) -- node[above, ylab] {yes} (val);

\def\dropY{2.1}

\coordinate (sno) at (-0.75,\dropY);
\draw[flow] (ds.south) |- (sno) |- (w1.west);
\node[nlab] at ($(ds.south)+(0.40,-0.30)$) {no};
\node[elabel] at ($(sno)+(5.1,0)$) {strength not proved};

\coordinate (wyes) at (dw.north|-0,\dropY);
\draw[flow] (dw.north) -- (wyes) -| (val.south);
\node[ylab] at ($(dw.north)+(0.42,0.30)$) {yes};

\draw[flow] (w6.north) -- node[elabel, pos=0.45] {still uncertified} (unout.south);
\draw[flow] (val) -- (out);

\begin{pgfonlayer}{background}
\node[lanebox, draw=cProof, fill=cProof!4, fit=(s1)(s2)(ds)] (stronglane) {};
\node[lanebox, draw=cWit, fill=cWit!4,
      fit=(w1)(w6)(note1)(note2)] (weaklane) {};
\end{pgfonlayer}
\node[lanelabel, text=cProof, anchor=west]
  at ($(stronglane.north west)+(4mm,0)$)
  {Certify Strong \textnormal{(\circnum{1}--\circnum{2})}\enspace
   \textnormal{\normalsize\itshape\textcolor{cMuted}{on every method}}};
\node[lanelabel, text=cWit, anchor=west]
  at ($(weaklane.north west)+(4mm,0)$)
  {Certify Weak \textnormal{(\circnum{3}--\circnum{8})}};
\end{tikzpicture}}

\caption{The \tool{} pipeline.
}
\label{fig:pipeline}
\end{figure}

\begin{figure}[t]
\lstset{basicstyle=\ttfamily\fontsize{5.8}{6.5}\selectfont}
\begin{minipage}[t]{0.485\textwidth}
\begin{lstlisting}[caption={A \texttt{proved\_strong} certificate for \datasetSmall's
\texttt{addPairwise} (task\_id\_470, RQ2-GPT4), which the benchmark correctly labels as strong.
},
label=lst:strong, style=certtarget]
method addPairwise(n: seq<int>) returns (res:seq<int>)
    requires n != []
    ensures |n| == 0 ==> |res| == 0
    ensures |n| > 0 ==> |res| == |n| - 1
    ensures forall i :: 
        0 <= i < |res| ==> res[i] == n[i] + n[i + 1]
{ 
    var result: seq<int> := []; 
    var i: int := 0;
    while i < |n| - 1
        // 3 invariants
    { 
        result := result + [n[i] + n[i + 1]]; 
        i := i + 1; 
    }
    res := result;
}
\end{lstlisting}
\certtag{cMuted}{$\blacktriangle$ target method $M$ and specification $S$, unchanged from the benchmark}
\certtag{cProof}{$\blacktriangledown$ certificate: the harness $H(S)$. \textbf{\texttt{dafny verify}} checks $M \models S$ and that $H(S)$ verifies}
\begin{lstlisting}[style=certpart]
///// FILLED IN HELPERS (empty) /////      // top level

// Predicate generated from addPairwise's `requires', verbatim.
ghost predicate addPairwise_Pre(n: seq<int>) 
{
    n != [] 
}

method addPairwise_check(n_original: seq<int>,
    n_mutant: seq<int>)
    returns (res_original: seq<int>, res_mutant: seq<int>)
    
    requires addPairwise_Pre(n_original)
    requires addPairwise_Pre(n_mutant)
    requires n_original == n_mutant      // inputs agree
    ensures res_original == res_mutant   // so must outputs
{
    ///// FILLED IN PROOF BEFORE ORIGINAL CALL (empty) /////
    res_original := addPairwise(n_original);
    ///// FILLED IN PROOF BEFORE MUTANT CALL (empty) /////
    res_mutant := addPairwise(n_mutant);
    ///// FILLED IN PROOF BEGIN (hints only; no assume) /////
    // FROM mode not-automatic
    assert |res_original| == |res_mutant|;
    forall i | 0 <= i < |res_original|
        ensures res_original[i] == res_mutant[i]
    {
        assert res_original[i] ==
            n_original[i] + n_original[i + 1];
        assert res_mutant[i] == n_mutant[i] + n_mutant[i + 1];
    } ///// FILLED IN PROOF END /////
}
\end{lstlisting}
\end{minipage}\hfill
\begin{minipage}[t]{0.485\textwidth}
\begin{lstlisting}[caption={A \texttt{witness\_weak} certificate for VeriEquivBench's \texttt{AddMinimumRungs} (\texttt{task\_id\_1936}, \emph{Medium}), which the benchmark labels as strong.},
label=lst:weak, style=certtarget]
method AddMinimumRungs(rungs: array<int>, dist: int)
    returns (result: int)
    requires rungs.Length >= 1  
    requires dist >= 1
    requires forall i :: 0 <= i < rungs.Length ==> rungs[i] >= 1
    requires forall i, j ::
        0 <= i < j < rungs.Length ==> rungs[i] < rungs[j]
    ensures result >= 0 // weak postcondition
{
    result := 0;  var currentHeight := 0;
    for i := 0 to rungs.Length  /* 5 invariants */ {
        var targetHeight := rungs[i];
        var gap := targetHeight - currentHeight;
        if gap > dist {
            var rungsNeeded := (gap - 1) / dist;
            result := result + rungsNeeded;  
        }
        currentHeight := targetHeight;
    }
}
\end{lstlisting}
\certtag{cMuted}{$\blacktriangle$ target method $M$ and specification $S$, unchanged from the benchmark}
\certtag{cWit}{$\blacktriangledown$ certificate: the mutant $M'$. \textbf{\texttt{dafny verify}} checks $M \models S$ and $M' \models S$}
\begin{lstlisting}[style=certpart]
///// FILLED IN HELPERS (empty) /////      // top level
method AddMinimumRungs_mutant(rungs: array<int>, dist: int)
    returns (result: int)  ... // regenerated, byte-compared
{   ///// FILLED IN COUNTEREXAMPLE BEGIN (no assume) /////
    // FROM mode dafnycert
    result := 0;
    ///// FILLED IN COUNTEREXAMPLE END /////
}
\end{lstlisting}
\certtag{cWit}{$\blacktriangledown$ certificate: the witness $x_0$. \textbf{\texttt{dafny run}} checks $x_0 \in X_P$ and $M(x_0) \neq M'(x_0)$}
\begin{lstlisting}[style=certpart]
method {:verify false} Main() {
    ///// FILLED IN WITNESS BEGIN (inputs only) /////
    // FROM mode dafnycert
    var rungs_original: array<int> := new int[] [463];
    var dist_original: int := 1;
    var rungs_mutant: array<int> := new int[] [463];
    var dist_mutant: int := 1;
    ///// FILLED IN WITNESS END /////
    expect rungs_original.Length >= 1; ...  // + 3 checks
    ///// FILLED IN PROOF BEFORE ORIGINAL CALL (empty) /////
    var originalResult :=
        AddMinimumRungs(rungs_original, dist_original);
    expect rungs_mutant.Length >= 1; ...    // + 3 checks
    ///// FILLED IN PROOF BEFORE MUTANT CALL (empty) /////
    var mutantResult :=
        AddMinimumRungs_mutant(rungs_mutant, dist_mutant);
    ///// FILLED IN PROOF AFTER CALLS (empty) /////
    print originalResult, " ", mutantResult;  // 462 and 0
    expect originalResult != mutantResult, "no divergence";
}
\end{lstlisting}
\end{minipage}
\end{figure}

\Cref{fig:pipeline} shows the \tool{} pipeline, which comprises three components. The \emph{Certify Strong} approach (\circnum{1}--\circnum{2}), the \emph{Certify Weak} approach (\circnum{3}--\circnum{8}), and the \emph{Certificate Validator} (\circnum{9}). The pipeline takes as input a method $M$ and its specification $S$, and outputs either a certificate of strength, such as the one in \Cref{lst:strong}, or a certificate of weakness, such as the one in \Cref{lst:weak}. In each, the shaded block is the target method, copied and unchanged from the benchmark, and every other block is the certificate, tagged with the Dafny command that checks it.

Both approaches are fully automatic. We first run Certify Strong on every method and Certify Weak on every method it does not certify. Every certificate, regardless of its source, must pass the validator, which detects tampering and unsound constructs. In \Cref{fig:pipeline}, each diamond marked \textsf{C} is a certificate candidate check. A candidate that passes it (\emph{yes}) goes to the validator, and one that fails it (\emph{no}) is discarded, moving the pipeline on to its next stage. Only after passing validation is the specification registered as weak or strong. If no certificate is found, the method is given a heuristic label using the mutation score (\circnum{8}).

\subsection{Definition of Weak and Strong Specifications}

\label{sec:background:semantics}
Informally, a specification is \emph{weak} if it accepts two distinct programs that disagree on some input, and \emph{strong} if it determines a unique behavior for every admissible input.
We assume throughout the paper that methods, denoted by \(M\), are \emph{deterministic}. This assumption is necessary for our notion of strength. Without it, a single implementation could produce different outputs on the same input, so differing outputs would not establish the existence of two distinct implementations. Consequently, randomized algorithms that produce nondeterministic outputs are outside the scope of this formalization.

Consider a method $M$ with input domain $X$ and output domain $Y$, where $Y$ represents both the returned values and the post-state of $M$'s \texttt{modifies} frame. The precondition is a
predicate $P : X \to \mathbb{B}$, the postcondition is a predicate
$Q : X \times Y \to \mathbb{B}$, and $S = (P,Q)$ is the
\emph{specification}. Dafny accepts $M$ only if it terminates whenever $P$
holds, so $M$ denotes a total function on the admissible inputs
$X_P = \{\, x \in X \mid P(x) \,\}$. Verifying $M$ against $S$, written $M \models S$, establishes that $Q(x,M(x))$ holds for all $x \in X_P$. We define strength through the set of outputs the specification admits for an input, $A_S(x) = \{\, y \in Y \mid Q(x,y) \,\}$.

\begin{definition}[Specification Strength]
\label{def:strength}
$S$ is \emph{strong} if $X_P \neq \emptyset$ and $|A_S(x)| = 1$ for every
$x \in X_P$, \emph{weak} if $|A_S(x)| > 1$ for some $x \in X_P$, \emph{vacuous} if $X_P = \emptyset$, and \emph{infeasible} if
$A_S(x) = \emptyset$ for some $x \in X_P$.
\end{definition}

\subsection{Certify Strong Approach}

The \emph{Certify Strong} approach checks \Cref{def:strength} indirectly, by verifying a harness. The harness takes two
copies $x_o$ and $x_m$ of the input, each required to satisfy $P$, calls $M$
once per copy to obtain $y_o$ and $y_m$, and states $y_o = y_m$ as its
postcondition.
For mutable inputs, the two copies have equivalent initial states but independent storage, so modifications made by the first call cannot affect the second. The comparison includes both returned values and the modified state.
Method \emph{addPairwise\_check} in \Cref{lst:strong} is an example of such a harness. 
Since Dafny reasons about a call only through the
callee's contract, the body of $M$ is opaque and the only available facts about
the two results are $Q(x_o,y_o)$ and $Q(x_m,y_m)$. Verifying the harness,
therefore, establishes that any two outputs the contract admits for one
admissible input coincide, i.e., $|A_S(x)| \le 1$. Written out, the obligation is:
\[
  \mathrm{VC}(S) \;\equiv\; \forall x_o, x_m \in X_P.\ \forall y_o, y_m \in Y.\
  x_o = x_m \wedge Q(x_o,y_o) \wedge Q(x_m,y_m) \implies y_o = y_m.
\]

\begin{definition}[Proof augmentation]
\label{def:augmentation}
A \emph{proof augmentation} $\Pi$ for $S$ consists of ghost Dafny code, which is used at verification time but never executed, yielding an augmented harness
$H(S,\Pi)$, with $H(S,\varepsilon) = H(S)$. $\Pi$ is \emph{admissible} when it
leaves $M$, $S$, and $H(S)$ untouched and introduces no  unproved claims.
\end{definition}

\begin{definition}[Certified strong]
\label{def:certstrong}
$S$ is \emph{certified strong} when some admissible $\Pi$, by default the
empty one, makes the solver report \texttt{unsat} for every obligation of
$H(S,\Pi)$, including $\neg\mathrm{VC}(S)$, and when, in addition, 
$\exists M.\, M \models S$ and $X_P \neq \emptyset$ hold.
\end{definition}

From \Cref{def:strength}, vacuity is ruled out by the side condition $X_P \neq \emptyset$, and infeasibility by the side condition $\exists M.\, M \models S$.
The certificate carries $M$ and verifies it, which establishes
$\exists M.\, M \models S$. 
To detect vacuity, the validator appends \texttt{assert false} at the end of the harness and re-verifies it. If Dafny proves it, the assumptions reaching that point are contradictory ($X_P = \emptyset$) and the certificate is rejected. Although this check detects vacuous specifications, its failure does not guarantee non-vacuity due to solver incompleteness. We therefore assume that specifications certified strong have satisfiable preconditions.

The augmentation $\Pi$ allows a specification to be
certified when its harness needs proof steps that the solver will not find on its
own, as is the case of ~\Cref{lst:strong}.

\begin{proposition}
\label{prop:certstrong}
If $S$ is certified strong then $S$ is strong.
\end{proposition}
\begin{proof}
Since $H(S,\Pi)$ verifies, its verification conditions $\mathrm{VC}(S)$ hold. Consider any $x \in X_P$, and instantiate the condition with $x_o = x_m = x$. For any $y_o, y_m \in A_S(x)$, the postcondition of the harness requires $y_o = y_m$. Hence $A_S(x)$ contains at most one element, i.e., $|A_S(x)|\leq 1$. The two side conditions guarantee that $A_S(x)$ is nonempty, and therefore $|A_S(x)|=1$. Thus $S$ is strong.
\end{proof}

\paragraph{Example.}
The method \texttt{addPairwise} from \Cref{lst:strong} computes the pairwise sums of consecutive elements in a sequence \texttt{n}. The specification completely states the expected length of the output sequence (empty when \texttt{n} is also empty and \texttt{|n| - 1} otherwise), and it also constrains the value of its elements (each element is the sum of the corresponding element in \texttt{n} and its immediate successor), fully capturing the method's intended behavior. 

The Certify Strong approach generates the harness (\circnum{1}) (\texttt{addPairwise\_check} in \Cref{lst:strong}). If it verifies (\circnum{2}), this shows that, for every admissible input, the method's specification admits exactly one output, making it strong. 
Dafny may, nonetheless, fail to verify the harness even when the specification is strong. This happens in \Cref{lst:strong}'s example because, although the specification establishes equality between corresponding elements of the two output sequences, Dafny does not automatically infer from these element-wise equalities that the sequences themselves are equal. The harness, therefore, needs a proof augmentation $\Pi$. The ghost \texttt{assert} and \texttt{forall} statements in the \texttt{PROOF} region of \Cref{lst:strong} establish sequence equality, enabling successful verification of the harness.

\subsection{Certify Weak Approach}
\label{sec:background:weakcert}

\begin{definition}[Certified weak]
\label{def:certweak}
$S$ is \emph{certified weak} by a pair $(M', x_0)$ when Dafny proves both
$M \models S$ and $M' \models S$, a runtime check of $P$ establishes that the witness $x_0$ is admissible ($x_0 \in X_P$), and running $M$ and $M'$ on $x_0$ produces different outputs.
\end{definition}

\Cref{lst:weak} exemplifies such a certificate. The mutant must verify ($M' \models S$), and running the file exhibits the input $x_0$ on which the mutant and the original program diverge. Preconditions are checked at runtime because arbitrary witnesses often cannot be statically verified to fulfill them. For example, Dafny may exhaust its budget trying to prove that a list of 100 elements is sorted, whereas the compiled check evaluates this directly. We exclude from the dataset methods for which the compiler cannot convert the preconditions into a runtime check. We informally rely on the soundness of Dafny's compiler and of its emitted precondition check (\S\ref{sec:threats}).

\begin{proposition}
\label{prop:certweak}
If $S$ is certified weak, then $S$ is weak.
\end{proposition}

\begin{proof}
The runtime check of P gives $x_0 \in X_P$. Verifying $M$ and $M'$ gives $Q(x_0, M(x_0))$
and $Q(x_0, M'(x_0))$, so both results lie in $A_S(x_0)$, and, by compiler
soundness, they correspond to the values produced by the runs, which differ. Hence
$|A_S(x_0)| > 1$. 
\end{proof}

\paragraph{Example}
In \Cref{lst:weak}, the method \texttt{AddMinimumRungs} determines the minimum number of rungs that must be added to a ladder so that its top can be reached from the floor, given the heights of its existing rungs in \texttt{rungs}, and so that no step to the top exceeds \texttt{dist}. Though this method is labeled as strong in \datasetBig, the method’s postcondition only requires \texttt{result} to be non-negative, and says nothing about how many rungs should be added. Our pipeline classifies this specification as weak, and the certificate in \Cref{lst:weak} was generated automatically by the Certify Weak approach.

The Certify Weak approach first compiles $M$ to Python and uses a frontend we developed for the \emph{Hypothesis}~\cite{Hypothesis} property-based testing library to sample inputs satisfying $M$'s precondition (\circnum{3}). For each sampled input, we record the output produced by the original implementation. We then run \emph{MutDafny}~\cite{Mutdafny} (\circnum{4}), a mutation tool for Dafny that generates mutants and runs the verifier on each of them. Mutants that verify are retained as \emph{alive mutants} (\circnum{5}).

To distinguish alternative implementations that are verified because the specification is weak from those that are simply equivalent to the original program, we execute both the original program and each alive mutant on the same sample of inputs and compare their outputs. The input selection and the execution of the original method take place prior to mutation (\circnum{3}), as this only needs to occur once. As MutDafny produces mutants (\circnum{4}), we run the alive ones (\circnum{5}) on the selected inputs and compare their outputs with those of the original method (\circnum{6}). If we find a mutant whose output differs from the original method’s, we build a candidate certificate and run Dafny on it (\circnum{7}). A divergence counts only once its certificate passes this check and, if it fails, the search moves on to the next divergent mutant. Once a certificate is successfully checked, we interrupt the mutation and alive mutant execution stages (\circnum{4}, \circnum{6}).

If every operator has been applied, or the time budget for mutation is exhausted without producing a certificate, the method remains uncertified. The search also yields a mutation score (\circnum{8}): the fraction of generated mutants that failed to verify, which serves as a heuristic for classifying specifications. A high mutation score suggests that the specification constrains the code well, so it is heuristically strong. Otherwise, it is heuristically weak.

In the certificate shown in \Cref{lst:weak}, the \texttt{FILLED IN COUNTEREXAMPLE} section records the mutant. It replaces the whole method body with the single assignment \texttt{result := 0}, an incorrect implementation that still verifies. The \texttt{FILLED IN WITNESS} section records the input on which the two outputs differ. In this case, the witness inputs were \texttt{rungs := new int[][463]} and \texttt{dist := 1}, for which the original method returns \texttt{462}, whereas the mutant returns \texttt{0}.

\subsection{Certificate Validator}
\label{sec:mot:validator}

The validator component is what allows a certificate to be trusted without trusting whoever produced it. Both Certify Strong and Certify Weak approaches produce \emph{candidate} certificates. The Certificate Validator (\circnum{9}) evaluates a certificate regardless of its source. Thus, a bug in a certificate producer (e.g., \tool, a human, or an LLM agent) can cause a certificate to be missed, but never an invalid one to be accepted.

A certificate \(C\) consists of the benchmark file plus generated code, in which only the \code{FILLED IN} marker regions may be written to. For a \emph{strong certificate}, the regions form \(\Pi\), which includes the top-level \code{HELPERS} region for lemmas and functions, and the proof-hint regions before each call to \(M\) and after both calls (in \Cref{lst:strong}, only the proof-hint regions are filled). For a \emph{weak certificate}, \code{COUNTEREXAMPLE} is the body of \(M'\) (\code{result := 0} in \Cref{lst:weak}), and \code{WITNESS} declares \(x_0\) once per call. The \code{HELPERS} and proof-hint regions in \code{Main} follow the same structure as for strong certificates. The listings show each empty region as a single \code{(empty)} line. Let $T(M,S)$ be the template regenerated from the benchmark source and $\mathit{blank}(C)$ be $C$ with its regions emptied. The validator accepts $C$ only if:

\begin{itemize}[leftmargin=*]
\item \textbf{V1: Structure.} The marker regions exist and are well-formed.
\item \textbf{V2: Grounding.} $\mathit{blank}(C)$ is byte-identical to $T(M,S)$, so $M$, $S$, $H(S)$, the contract of $M'$, and the checks in \texttt{Main} are unchanged (i.e., in \Cref{lst:strong,lst:weak}, everything outside the \texttt{FILLED IN} blocks).
\item \textbf{V3: No unsound constructs.} No \texttt{assume} appears in a strong certificate, or in a weak certificate's regions, and no region declares a lemma, method, or function taken on trust (\texttt{\{:axiom\}}, \texttt{\{:verify false\}}, or no body). Hence $\Pi$ is admissible (\Cref{def:augmentation}) and $M'$ cannot verify by assumption. Weak regions also contain no \texttt{expect}, which could alter the run or skip a precondition check.
\item \textbf{V4: Verification.} \texttt{dafny verify} succeeds, establishing $M \models S$ and $H(S,\Pi)$ for a strong certificate, and $M \models S$ and $M' \models S$ for a weak one.
\item \textbf{V5: Non-vacuity} (strong only). The harness with \texttt{assert false} appended does \emph{not} verify (see \Cref{def:certstrong}).
\item \textbf{V6: Same input} (weak only). $M$ and $M'$ receive inputs with equivalent initial states, using independent storage for mutable values. (\texttt{[463]} and \texttt{1} in \Cref{lst:weak}).
\item \textbf{V7: Divergence} (weak only). \texttt{dafny run} passes every runtime check of $P$, so $x_0 \in X_P$, and confirms $M(x_0) \neq M'(x_0)$ (\texttt{462} vs.\ \texttt{0} in \Cref{lst:weak}).
\end{itemize}

The validator checks V1--V3 and V6 itself and delegates V4, V5, and V7 to Dafny. 
Under the non-vacuity assumption, V2--V4 establish \Cref{def:certstrong}, while V5 provides an additional check
for vacuity.

V2, V4, V6, and V7 discharge \Cref{def:certweak}, so by \Cref{prop:certstrong,prop:certweak} an accepted certificate implies its label. The trusted base is thus Dafny (its verifier and, for weak certificates, its compiler) and the validator (\Cref{sec:threats}).

\section{Automatically Certifying Strong Specifications}
\label{sec:arch:strong}

The certification of strong specifications (\circnum{1}--\circnum{2}) runs first on every method and tries to show that every implementation of $S$ agrees with $M$ on every admissible input, which it does by generating a harness $H(S)$ (\circnum{1}) that Dafny itself verifies (\circnum{2}). If verification fails, the method proceeds to the certification of weak specifications.

The main challenge is constructing a harness such that Dafny has enough information to verify it automatically. Our design builds on the equivalence check introduced by \datasetBig{}~\cite{VeriEquivBench}, which we call \veqcheck{}, but uses a different encoding and supports a broader range of Dafny constructs.

\subsection{A Robust Equivalence Harness}
\label{sec:arch:strong:veriequiv}

\veqcheck{} imposes a similar obligation with a single call. It havocs the outputs, assumes $P$ and $Q$ on the havoced values, calls $M$ once, and asserts that the two agree. The logical content is the same, the difference is how many facts about the outputs must be restated by hand. A havoced output does not retain several facts that a normally returned value carries automatically, such as non-nullness or prior allocation, and the \veqcheck{} encoding does not restate all of them. 

This difference changes verification outcomes. 
For \datasetSmall{}'s \code{SquareArrayElements} method (task\_id\_8, RQ1-GPT4), \veqcheck{}'s generator would produce the following harness:

\begin{lstlisting}[basicstyle=\ttfamily\fontsize{6.2}{7.0}\selectfont]
method SquareArrayElements_check(arr: array<int>) returns (res: array<int>) {
  res := *;
  assume arr != null;
  assume res != null;
  assume arr.Length == res.Length;
  assume forall i :: 0 <= i < arr.Length ==> res[i] == arr[i] * arr[i];
  var val_0 := SquareArrayElements(arr);
  assert res.Length == val_0.Length;
  assert forall i :: 0 <= i < res.Length ==> res[i] == val_0[i]; // 1 error
}
\end{lstlisting}
Despite the specification being strong, this harness does not verify. Since \texttt{res} is havoced, Dafny cannot establish that it was allocated before the call. Consequently, the callee's frame does not preserve any information about it, and the assumptions on its contents are lost. Adding an \texttt{assume allocated(res)} would be enough to restore the missing allocation information and allow the harness to verify. In contracts without an explicit \texttt{ensures res != null}, havocing also drops the declared type's non-nullness, leaving \texttt{res.Length} and \texttt{res[i]} ill-formed.

Our harness avoids these problems by making two actual calls to the method, so $Q$ reaches the check through the callee's contract, and each result keeps its type, allocation, and frame information. The harness is also regular Dafny code without explicit \texttt{assumes}, which makes it easier to re-check. 
For methods that modify the heap, additional proof steps may be needed to establish the equivalence of the modified states.

\subsection{Supporting a Broader Range of Dafny Programs}

Beyond the harness encoding, we developed a generator that supports a broader range of Dafny programs. \veqcheck{}'s generator recovers method signatures using regular expressions, which can fail on nested parentheses and on identifiers that start with a keyword. Our generator instead extracts the method header from Dafny's resolved AST.
We also extend support for several Dafny constructs. \veqcheck{} hardcodes one-dimensional array accessors, treats the value type \code{map} as mutable, and does not resolve \code{type} synonyms. Our generator handles array rank explicitly, distinguishes value and reference types, and resolves type synonyms. \veqcheck{} also omits the equality check for an array that is both returned and modified, which can cause a weak specification to be classified as strong, although this did not occur in our evaluation.  Finally, it does not model the receiver of an instance method, so modifications to receiver fields are not compared. Our generator explicitly represents the receiver and its \code{modifies} frame.
\Cref{sec:rq1} compares the certification rates of \tool{} and \veqcheck{}.

\section{Automatically Certifying Weak Specifications}
\label{sec:arch:weak}

Certifying a specification as weak requires finding an alternative implementation $M'$ that verifies against the original specification $S$, together with an admissible input $x_0$ on which $M$ and $M'$ produce different outputs. To certify weak specifications, \tool{} searches for $M'$ using program mutations and for $x_0$ using property-based input generation (\circnum{3}--\circnum{8}).

Two challenges arise. First, the search must find a mutant that both verifies and behaves differently, but MutDafny's operators were not designed to expose specification weaknesses. Second, the search must be affordable. A method may admit hundreds of mutants, each of which must be verified, and restrictive preconditions make admissible inputs hard to generate. We address the first challenge with mutation operators aimed at specification weaknesses (\Cref{sec:search:mutants}). We address the second with input generation guided by the method's preconditions (\Cref{sec:search:witness}) and with a reduced set of mutation operators that keeps most of the witnesses while evaluating far fewer mutants (\Cref{sec:restricting_mutant_set}).

\subsection{Generating Alternative Implementations}
\label{sec:search:mutants}

We use MutDafny~\cite{Mutdafny} to generate candidate alternative implementations. This tool applies syntactic mutation operators to the method
under test, generating one implementation $M'$ per mutation (\circnum{4}), and verifies each against the original specification $S$. A mutant that verifies is therefore another implementation of $S$ and is retained as an \emph{alive mutant} (\circnum{5}). However, an alive mutant may still be behaviorally equivalent to the original implementation $M$, so verification alone is not sufficient to establish weakness. In \Cref{sec:search:witness}, we search for an input that distinguishes $M$ and $M'$.

MutDafny defines 40 mutation operators, most being local perturbations of existing code. Thirty are adapted from operators implemented in mutation tools for other programming languages, while the remaining ten were derived from an analysis of implementation mistakes made by Dafny developers. To find the kinds of specification weakness MutDafny misses, two of the authors, each with three years of experience with Dafny, inspected the \datasetSmall{} methods for which no alive mutant was found. For each method, they identified the behavior the specification leaves unconstrained and designed one operator whose mutant changes
that behavior while still verifying. This yields the nine operators of \Cref{tab:added_operators}, which target collection content, collection order, arithmetic expressions, and the method as a whole.

\begin{table}[t]
\caption{The nine mutation operators we add to MutDafny's catalogue}
\label{tab:added_operators}
\footnotesize
\newcommand{\opgroup}[3]{\midrule\multicolumn{3}{@{}p{\dimexpr\textwidth}@{}}{\textbf{#1.} \emph{Weakness:} #2 \emph{Gap:} #3}\\[1pt]}
\begin{tabularx}{\textwidth}{@{}l l X@{}}
\toprule
Op. & Name & Mutation \\
\opgroup{Collection content}{a collection's length is constrained, not its elements.}{only initializers change (CIR).}
CEU & Collection Element Update & Sets the first or last element of a collection to a default value \\
CCC & Collection Content Copy & Assigns the contents of another collection to a collection \\
CCU & Collection Comprehension Update & Assigns a default comprehension expression to a collection \\
\opgroup{Collection order}{the elements are constrained, but not their order.}{no operator targets element order.}
CES & Collection Element Swap & Swaps the first two elements of an ordered collection \\
CAS & Sequence Concatenation Argument Swap & Swaps the two operands of a sequence concatenation \\
\opgroup{Arithmetic}{loose bounds, off-by-one results, unconstrained lengths.}{LVR mutates only literals.}
INC, DEC & Arithmetic Expression Increment/Decrement & Replaces a non-literal arithmetic expression $e$ with $e \pm 1$ \\
ELR & Expression-Collection Length Replacement & Replaces an arithmetic expression with a collection's length \\
\opgroup{Whole method}{a trivial output satisfies $Q$ (\Cref{lst:weak}).}{all operators are local.}
VSR & Vacuous Stub Replacement & Replaces the body with default assignments to the outputs \\
\bottomrule
\end{tabularx}
\end{table}

\subsection{Generating Witness Inputs}
\label{sec:search:witness}

Given $M$ and an alive mutant $M'$, we need to find an $x_0 \in X_P$ for which $M(x_0) \neq M'(x_0)$. We do this by running both methods on the same sample of inputs and comparing their outputs.

The sample of inputs is generated using a frontend we built for the \emph{Hypothesis}~\cite{Hypothesis} property-based testing library for Python. To this end, both programs are compiled to Python by running \texttt{dafny build --target:py}. To ensure that candidate inputs satisfy $M$'s precondition \(P\), we also compile a \emph{checked} variant using Dafny's \texttt{--test-assumptions Externs} option, which raises an exception whenever a precondition is violated, so that every accepted input satisfies \(P\). The inputs are drawn once against $M$ (\circnum{3}), and $M$ and $M'$ are both executed on them and their results compared (\circnum{6}). The sampler targets \numAblationMaxExamples{} admissible inputs per method.

Our sampling pipeline has two stages. First, our Hypothesis frontend maps each Dafny type to a Hypothesis strategy, so that candidate inputs can be drawn for every supported type. 
A strategy exists for the primitive types bool, nat, int, real,
char, and string. Collections, including seq, set, map, and up-to-four-dimensional arrays, are
supported recursively over their element types. Tuples of any arity and nesting are also supported,
as well as generic and directly recursive datatypes. Support is currently missing for multiset,
recursive classes and traits, and mutually recursive datatypes.

The frontend's reliance on filtering (i.e., drawing candidate inputs without regard to $P$ and discarding those that the $P$ runtime check rejects) works well for relatively permissive preconditions, such as scalar bounds, but is ineffective for restrictive ones. For instance, an unguided draw has a very low probability of ever producing a strictly increasing array of positive integers, making it very difficult to find admissible inputs for methods with such a precondition and, as a result, to find a witness for them.
To address this, a second component parses and identifies common precondition patterns in Dafny's resolved AST and emits a Hypothesis composite strategy that incorporates these constraints when generating inputs, rather than filtering them out afterward. Recognized constraints are satisfied by construction, while all other cases fall back to filtering.
Across the \numPrecondMethods{} methods with preconditions in the dataset, this component recognizes \numPrecondConjunctsPct{}\% of the \numPrecondConjuncts{} recorded precondition elements. Of those methods, \numPrecondFullPct{}\% have every precondition element encoded into the sampling strategy, \numPrecondPartPct{}\% have at least one, and only \numPrecondNoneMethods{} are left entirely to filtering.

\Cref{tab:recognise} lists the supported precondition patterns and the Hypothesis constraints instantiated for each, which are then assembled into a \texttt{@st.composite} strategy.
Parameters are drawn in an order that ensures every symbolic bound refers only to parameters that have already been drawn. 

The recognized constraints are those that can be built directly: bounds, memberships, distinctness, sortedness, row widths, column bounds, and calls to inlined helper predicates. Existential quantifiers, predicates that cannot be inlined, and recursive predicates remain out of scope.

\begin{table}[t]
\caption{The matched Dafny precondition clause patterns and how each of those patterns is converted into a Hypothesis strategy.
\texttt{p} and \texttt{q} are parameters, \texttt{k} a literal, \texttt{w} a literal row width, \texttt{a}, \texttt{b} and \texttt{c} literal column indices, $\bowtie$ a comparison, \texttt{f(q)} an expression over already-drawn parameters, and \texttt{lo} and \texttt{hi} the accumulated minimum and maximum of same-direction bounds.}\label{tab:recognise}
\small
\begin{tabularx}{\textwidth}{@{}>{\raggedright\arraybackslash\hsize=1\hsize}X>{\raggedright\arraybackslash\hsize=1\hsize}X@{}}
\toprule
Precondition Clause Shape & Hypothesis Constraint Rule \\
\midrule
\footnotesize\texttt{p $\bowtie k$}, \texttt{p $\bowtie$ f(q)}

  & \footnotesize\texttt{st.integers(min\_value=lo, max\_value=hi)} \\
  \midrule
\footnotesize\texttt{p == a$_1$ || ... || p == a$_m$}

  & \footnotesize\texttt{st.sampled\_from([a$_1$,..,a$_m$])} \\
  \midrule
\footnotesize\texttt{|p| $\bowtie$ k, |p| $\bowtie$ f(q)}

  & \footnotesize\texttt{st.lists(min\_size=lo, max\_size=hi)} \\
  \midrule
\footnotesize\texttt{forall i, j :: 0 <= i < j < |p| ==> p[i] != p[j]}

  & \footnotesize\texttt{st.lists(unique=True)} \\
  \midrule
\footnotesize\texttt{forall i :: p[i] $\bowtie$ p[i+1]}, \newline
\footnotesize\texttt{forall i, j :: i < j ==> p[i] $\bowtie$ p[j]}

  & \footnotesize\texttt{st.lists(...).map(sorted(reverse=rev))}, \footnotesize\texttt{rev} \small is \footnotesize\texttt{True} \small or \footnotesize\texttt{False} \small depending on the $\bowtie$ direction \\
  \midrule
\footnotesize\texttt{forall i :: 0 <= i < |p| ==> p[i] $\bowtie$ k}
   & for each \footnotesize\texttt{p[i]}, \footnotesize\texttt{p[i] = st.integers(min\_value=lo, max\_value=hi)} \\
  \midrule
\footnotesize\texttt{forall i :: 0 <= i < |p| ==> \newline \hspace*{2em}p[i] == a$_1$ || ... || p[i] == a$_m$}

  & for each \footnotesize\texttt{p[i]}, \footnotesize\texttt{p[i] = st.sampled\_\-from([a$_1$,..,a$_m$])} \\
  \midrule
\footnotesize\texttt{p[k] == f(q)}

  & generates \footnotesize\texttt{p}\small, then sets \footnotesize\texttt{p[k] = f(q)} \\
  \midrule
\footnotesize\texttt{forall i :: 0 <= i < |p| ==> |p[i]| == w}

  & \footnotesize\texttt{st.lists(st.tuples(st.integers(), st.integers(),} \newline
    \footnotesize\texttt{\hspace*{4.5em}st.integers()).map(list),} \newline
    \footnotesize\texttt{\hspace*{1.2em}min\_size=lo, max\_size=hi)\hfill\# w == 3} \\

  \midrule
\footnotesize\texttt{forall i :: 0 <= i < |p| ==> p[i][c] $\bowtie$ k}

  & \footnotesize\texttt{st.tuples(st.integers(),} \newline
    \footnotesize\texttt{\hspace*{1.2em}st.integers(min\_value=lo, max\_value=hi),} \newline
    \footnotesize\texttt{\hspace*{1.2em}st.integers()).map(list)\hfill\# c == 1} \\

  \midrule
\footnotesize\texttt{forall i, j :: p[i][j] $\bowtie$ k}

  & \footnotesize\texttt{st.lists(st.lists(} \newline
    \footnotesize\texttt{\hspace*{1.2em}st.integers(min\_value=lo, max\_value=hi)))} \newline
    \footnotesize\texttt{\# one st.lists per dimension} \\
\bottomrule
\end{tabularx}
\end{table}

\Cref{lst:weak}'s preconditions require \texttt{rungs} to be a strictly increasing array of positive integers with at least one element, and \texttt{dist} to be a positive integer. If we blindly draw values and filter afterward, such an array almost never satisfies the precondition. The frontend instead emits a strategy that builds it directly:

\begin{lstlisting}[language=Python,basicstyle=\ttfamily\fontsize{7.5}{9}\selectfont]
@st.composite
def addminimumrungs_precondition_strategy(draw):
    rungs = draw(st.lists(st.integers(min_value=1),
                          min_size=1, unique=True).map(sorted))
    dist = draw(st.integers(min_value=1))
    return (rungs, dist)
\end{lstlisting}

A successful witness search (\circnum{6}) returns an input the checked variant accepted, and the certificate records it so that anyone can reproduce the divergence. The certificate also emits $P$ as a check before each call, so running it re-tests admissibility.

As \Cref{sec:background:weakcert} describes, a runtime disagreement becomes a certificate only once the candidate file passes \texttt{dafny verify} and \texttt{dafny run}. Otherwise, the search moves on to the next divergent mutant, and a method with no certificate keeps only its mutation score (\circnum{8}).

\subsection{Reducing the Mutation Search Space}
\label{sec:restricting_mutant_set}

Generating and verifying every mutant is expensive~\cite{Mutdafny}. Moreover, mutation operators often overlap in the weaknesses they expose. We therefore select a smaller set of operators that preserves witness-finding coverage while minimizing the number of mutants that must be verified.

For each operator, we define its \emph{coverage set} as the set of methods for which at least one mutant generated by it both verifies and yields a witness. We use these coverage sets to select a minimum-cost subset of operators by formulating the selection as a \emph{minimum-cost partial set cover}~\cite{PartialCover}, which minimizes cost under a coverage floor. We solve this optimization over a sample of our dataset, and then apply the resulting subset to the full dataset.

For each operator $o$, we introduce one binary decision variable $z_o$, set to $1$ when operator $o$ is kept and $0$ when it is dropped, and define $C_o$ as the total number of mutants generated by $o$ across the sample, i.e., the cost of keeping $o$. The coverage sets are represented by $E_{om} \in \{0,1\}$, which, for each method $m$ in our sample that is covered by the full operator set, records whether operator $o$ is capable of producing an alive mutant and witness for $m$. For each method $m$, we also introduce a binary variable $O_m \in \{0,1\}$ to indicate whether it is covered by the selected operators. Our goal is to minimize the total number of mutants generated by the kept operators, $\sum_o C_o z_o$, while achieving a coverage level of $p$ among the $N$ methods covered by the full operator set:

\[
  \min \sum_o C_o z_o
  \quad \text{s.t.} \quad
  O_m \le \sum_{o\,:\,E_{om}=1} z_o \ \ \forall m,
  \qquad
  \sum_m O_m \ge p \cdot N,
  \qquad
  \text{$O_m, z_o \in \{0,1\}$},
\]

The first constraint forces \(O_m = 0\) when no kept operator covers method \(m\). 
Since the solver may also set \(O_m = 0\) for a covered method once the floor is met, we compute the reported coverage from the kept operators \(z_o\) rather than from \(\sum_m O_m\). This does not change the optimum.

Setting \(p=1\) requires the selected operators to cover every method in the sample, which corresponds to the standard \emph{minimum-cost set cover} problem \cite{SetCover}. As an ablation, we sweep \(p\) from 0.5 to 1 (\Cref{sec:rq3}). Because \(p\) is only a floor, a solution may cover more methods than it requires, so ~\Cref{tab:operators} reports both the target \(p\) and the coverage obtained.

To measure the coverage sets $E_{om}$, we must evaluate all operators rather than stop at the first certificate. We therefore run every operator on each method in the \textsf{Selection Set}, without early stopping and with an extended per-method budget. The \textsf{Selection Set} is a random sample of \numSampleTotal{} \datasetBig{} methods (\Cref{sec:rq3}). With this data, operator selection is performed and the resulting operator set is evaluated on methods that the selection never saw (\textsf{Transfer Set}). We deploy the set selected at $p=1$.

\section{Evaluation}

We evaluate \tool{} on a dataset of \numCorpusTotal{} Dafny methods from Dafny-Synthesis~\cite{dafny_synthesys_FSE24} and Ve\-ri\-EquivBench~\cite{VeriEquivBench}. Our evaluation examines the effectiveness of automatic certification, the agreement between certified and existing benchmark labels, and the trade-offs involved in generating weakness certificates.
We organize our evaluation around the following research questions:
\begin{description}
    \item[\textbf{RQ1:}] How effectively does \tool{} automatically certify specifications as strong or weak?
    \item[\textbf{RQ2:}] How do the certified labels produced by \tool{} compare with those provided by existing benchmarks?
    \item[\textbf{RQ3:}] How do the strategies used to generate weakness certificates affect certification effectiveness and computational cost?
\end{description}

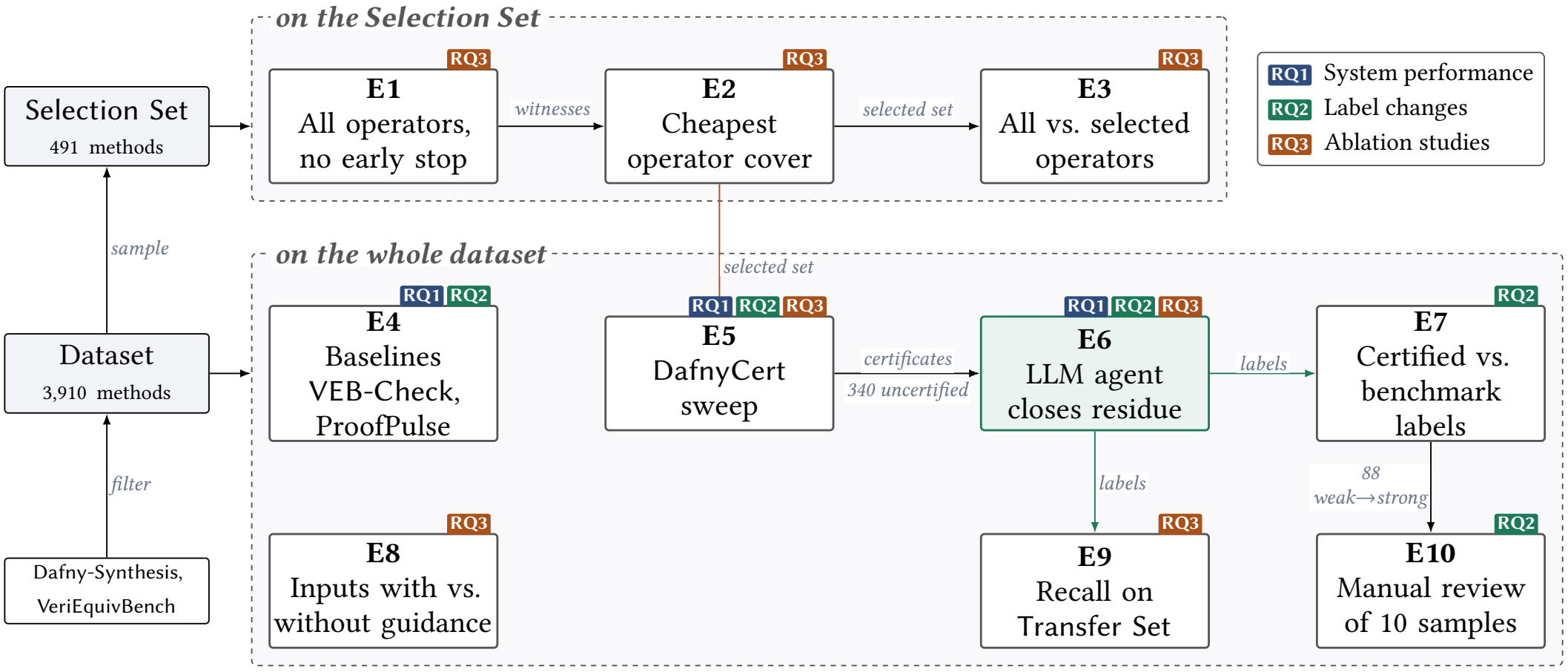
\begin{figure}[t]
\centering
\resizebox{\linewidth}{!}{%
\hyphenpenalty=10000\exhyphenpenalty=10000%
\begin{tikzpicture}[
  font=\Large,
  expbox/.style={rectangle, rounded corners=2pt, draw=cValid, very thick,
    align=center, text width=4.4cm, minimum height=2.3cm, inner sep=3pt,
    fill=white, font=\huge},
  truthbox/.style={expbox, draw=cProof, fill=cProof!9},
  iobox/.style={rectangle, rounded corners=2pt, draw=black, thick,
    align=center, text width=3.9cm, minimum height=1.6cm, fill=cRule!25, font=\huge},
  badge/.style={rectangle, rounded corners=1.5pt, inner sep=1.6pt,
    font=\large\bfseries\sffamily, text=white},
  rqone/.style={badge, fill=cAgent},
  rqtwo/.style={badge, fill=cProof},
  rqthree/.style={badge, fill=cWit},
  flow/.style={-{Latex[length=2.2mm]}, thick},
  elabel/.style={font=\Large\itshape, fill=white, inner sep=1.5pt,
    text=cMuted, align=center},
  lanelabel/.style={font=\huge\bfseries\itshape, fill=white, inner sep=2pt},
  lanebox/.style={rounded corners=3pt, dashed, thick, inner sep=3.4mm},
]
\def\cA{5.2}\def\cB{12.0}\def\cC{19.6}\def\cD{26.4}
\def\rTop{0}\def\rMid{-5.0}\def\rLow{-9.4}

\node[expbox] (allops) at (\cA,\rTop)
  {\textbf{E1}\\All operators,\\no early stop};
\node[expbox] (cover) at (\cB,\rTop)
  {\textbf{E2}\\Cheapest\\operator cover};
\node[expbox] (ab) at (\cC,\rTop)
  {\textbf{E3}\\All vs.\ selected\\operators};
\draw[flow] (allops) -- node[elabel, above=1.5mm] {witnesses} (cover);
\draw[flow] (cover) -- node[elabel, above=1.5mm] {selected set} (ab);

\node[expbox] (base) at (\cA,\rMid)
  {\textbf{E4}\\Baselines\\\veqcheck{}, ProofPulse};
\node[expbox] (sweep) at (\cB,\rMid)
  {\textbf{E5}\\\tool{} sweep};
\node[truthbox] (truth) at (\cC,\rMid)
  {\textbf{E6}\\LLM agent\\closes residue};
\node[expbox] (cmp) at (\cD,\rMid)
  {\textbf{E7}\\Certified vs.\\benchmark labels};
\node[expbox] (guide) at (\cA,\rLow)
  {\textbf{E8}\\Inputs with vs.\\without guidance};
\node[expbox] (held) at (\cC,\rLow)
  {\textbf{E9}\\Recall on\\\textsf{Transfer Set}};
\node[expbox] (review) at (\cD,\rLow)
  {\textbf{E10}\\Manual review\\of \numWeakToStrongReviewed{} samples};

\draw[flow, cWit] (cover) -- node[elabel, right, pos=0.62] {selected set} (sweep);
\draw[flow] (sweep) -- node[elabel, above] {certificates}
  node[elabel, below=1mm]
  {\numUncertified{} uncertified} (truth);
\draw[flow, cProof] (truth) -- node[elabel, above] {labels} (cmp);
\draw[flow, cProof] (truth) -- node[elabel, right] {labels} (held);
\draw[flow] (cmp) -- node[elabel, left] {\numVbSingleToStrong{}\\weak$\rightarrow$strong} (review);

\newcommand{\badges}[2]{%
  \node[anchor=south east, inner sep=0pt] at ($(#1.north east)+(-1.5mm,-0.4mm)$)
    {#2};}
\newcommand{\bOne}{\tikz[baseline]\node[rqone, anchor=base]{RQ1};}
\newcommand{\bTwo}{\tikz[baseline]\node[rqtwo, anchor=base]{RQ2};}
\newcommand{\bThree}{\tikz[baseline]\node[rqthree, anchor=base]{RQ3};}
\badges{allops}{\bThree}
\badges{cover}{\bThree}
\badges{ab}{\bThree}
\badges{base}{\bOne\,\bTwo}
\badges{sweep}{\bOne\,\bTwo\,\bThree}
\badges{truth}{\bOne\,\bTwo\,\bThree}
\badges{cmp}{\bTwo}
\badges{guide}{\bThree}
\badges{held}{\bThree}
\badges{review}{\bTwo}

\node[iobox] (sel) at (-0.4,\rTop) {\textsf{Selection Set}\\{\Large\numSampleTotal{} methods}};
\node[iobox] (corpus) at (-0.4,\rMid) {Dataset\\{\Large\numCorpusTotal{} methods}};
\node[iobox, minimum height=1.0cm, fill=white] (src) at (-0.4,\rLow)
  {{\Large\datasetSmall{},\\\datasetBig{}}};
\draw[flow] (src) -- node[elabel, right] {filter} (corpus);
\draw[flow] (corpus) -- node[elabel, right] {sample} (sel);

\coordinate (selTop) at ($(allops.north)+(0,7mm)$);
\coordinate (corTop) at ($(base.north)+(0,7mm)$);
\begin{pgfonlayer}{background}
\node[lanebox, draw=cValid, fill=cRule!12, fit=(allops)(ab)(selTop)] (sellane) {};
\node[lanebox, draw=cValid, fill=cRule!12,
      fit=(base)(review)(guide)(held)(corTop)] (corlane) {};
\end{pgfonlayer}
\node[lanelabel, text=cValid, anchor=west] at ($(sellane.north west)+(4mm,0)$)
  {on the \textsf{Selection Set}};
\node[lanelabel, text=cValid, anchor=west] at ($(corlane.north west)+(4mm,0)$)
  {on the whole dataset};
\draw[flow] (sel) -- (sel-|sellane.west);
\draw[flow] (corpus) -- (corpus-|corlane.west);

\node[rectangle, rounded corners=3pt, draw=cValid, thick, fill=white,
      align=left, anchor=west, inner sep=6pt, font=\LARGE]
  at ($(sellane.east)+(6mm,0)$)
  {\bOne\enspace System performance\\[3pt]
   \bTwo\enspace Label changes\\[3pt]
   \bThree\enspace Ablation studies};
\end{tikzpicture}}
\caption{The experiments E1--E10 of the evaluation.}
\label{fig:evaluation}
\end{figure}

\Cref{fig:evaluation} shows the experiments we run, the data each one runs on, and the research questions that use its results. Arrows show what one experiment passes to the next. The dataset run E5 uses the operator set that E2 selects, and the methods it leaves uncertified go to E6. Each paragraph heading names the experiments it reports.

\subsection{Dataset Construction and Experimental Setup}
\label{sec:dataset}

\subsubsection*{Datasets.}
Dafny-Synthesis~\cite{dafny_synthesys_FSE24} synthesizes Dafny methods for MBPP-derived programming tasks using GPT-4 and PaLM-2 under several prompting strategies. Each file contains a single method, and its specification is labeled by human annotators according to whether it captures the intended behavior of the corresponding natural-language task.

VeriEquivBench~\cite{VeriEquivBench} translates LeetCode problems into Dafny and may include multiple methods in a single file. Its labels are assigned at the file level, with an LLM prompted to generate one strong and one weak specification for each problem. The corresponding label therefore applies to every method in the file.

\subsubsection*{Labels.}
The labels provided by both benchmarks are \emph{extrinsic}, in the sense that they assess whether a specification captures the intended behavior described by the corresponding natural-language task. In contrast, our certified labels are \emph{intrinsic}, since they depend only on the specification itself and characterize whether it admits more than one observable behavior, independently of the natural-language description.

These notions are related but not equivalent. For tasks with a uniquely determined intended behavior, a specification that we certify as weak cannot fully capture that behavior, since its certificate exhibits two implementations that satisfy the same specification but produce different outputs on the same admissible input. Conversely, a specification that we certify as strong may still uniquely characterize behavior that does not match the natural-language intent. We account for this distinction when analyzing disagreements with the benchmark labels in RQ2.

\subsubsection*{Dataset construction.}
We restrict the dataset to methods for which certificates can potentially be generated and validated. Overall, this filtering excludes \numExcludedTotal{} methods, with by far the largest category consisting of \numExcludedFileVerify{} methods which are not verified by Dafny. The remaining exclusions include files that do not compile, specifications containing untrusted constructs, methods without observable behavior, unsupported self-referential reference types, specifications labeled \emph{wrong} in Dafny-Synthesis because their implementations are unrelated to the natural-language descriptions, and a small number of cases that could not be executed within the available resources. The resulting dataset contains \numCorpusTotal{} methods across \numCorpusFiles{} files, including \numVeribenchTotal{} methods from VeriEquivBench and \numSynthTotal{} from Dafny-Synthesis.
After completing the certification process, \numCorpusStrong{} specifications are certified strong and \numCorpusWeak{} are certified weak, as summarized in Table~\ref{tab:classification}. The table also shows the \textsf{Selection Set}, the subset of \datasetBig{} used to select mutation operators, and the \textsf{Transfer Set}, the rest of \datasetBig{}, on which we measure how that selection generalizes (\Cref{sec:rq3}).

\begin{table}[t]
\begingroup
\renewenvironment{table}[1][]{\begin{minipage}[t]{\subtablewidth}\setlength{\columnwidth}{\linewidth}\centering}{\end{minipage}}
\def\subtablewidth{0.37\linewidth}\begin{table}[t]
\caption{Dataset certified classification.}
\label{tab:classification}
\footnotesize
\setlength{\tabcolsep}{4pt}
\begin{tabular*}{\columnwidth}{@{\extracolsep{\fill}}lrrr@{}}
\toprule
& Methods & Strong & Weak \\
\midrule
\datasetBig{} & 3{,}692 & 420 & 3{,}272 \\
\quad\textsf{Selection Set} & 491 & 60 & 431 \\
\quad\textsf{Transfer Set} & 3{,}201 & 360 & 2{,}841 \\
\datasetSmall{} & 218 & 167 & 51 \\
\midrule
Dataset & 3{,}910 & 587 & 3{,}323 \\
\bottomrule
\end{tabular*}
\end{table}
\hfill
\def\subtablewidth{0.6\linewidth}\begin{table}[t]
\centering
\caption{Relabellings by direction.}
\label{tab:flips}
\footnotesize
\setlength{\tabcolsep}{3pt}
\begin{tabular}{@{}lrrr@{}}
\toprule
Population & str.$\rightarrow$weak & weak$\rightarrow$str. & changed \\
\midrule
\datasetBig{} (all)          & \numVeribenchToWeak{}  & \numVeribenchToStrong{}  & \numVeribenchChanged{} \\
\quad single-method files    & \numVbSingleToWeak{}   & \numVbSingleToStrong{}   & \numVbSingleChanged{} \\
\datasetSmall{}              & \numSynthToWeak{}      & \numSynthToStrong{}      & \numSynthChanged{} \\
\bottomrule
\end{tabular}
\end{table}

\endgroup
\end{table}

\subsubsection*{Experimental Setup}
\label{sec:eval:setup}

All experiments were conducted on a workstation with a 16-core Intel Core Ultra 9 285H processor and 30~GiB of RAM, running Linux 6.18. We use Dafny~4.11.1 with Z3~4.12.1 inside the project's container image. Experiments are run with up to~\numToolConcurrency{} methods in parallel. Unless otherwise stated, reported execution times correspond to the sum of the elapsed time across individual methods rather than wall-clock time. Because the weakness search has a time budget per method and generates its inputs randomly, rerunning DafnyCert may produce slightly different results.
The harness and validator have no wall-clock limit and rely only on Z3's deterministic resource limit (\numBudgetResourceMillions{}M units per assertion batch, the same for the harness, the vacuity probe, and the validator). The weakness search has a budget of \numBudgetMutationSeconds{}\,s per method, with only \numToolTimeout{} of the \numCorpusTotal{} methods timing out.

\subsection{RQ1: How effectively does \tool{} automatically certify specifications as strong or weak?}
\label{sec:rq1}

\begin{table*}[t]
\caption{Comparison between approaches that classify the strength of specifications over the certified dataset.}
\label{tab:rq1matrix}
\footnotesize
\setlength{\tabcolsep}{4pt}
\begin{tabular*}{\columnwidth}{@{\extracolsep{\fill}}lrrrrrrrrcrrr@{}}
\toprule
& \multicolumn{2}{c}{weak} & \multicolumn{2}{c}{strong} & \multicolumn{5}{c}{classified} & \multicolumn{2}{c}{MCC} & \\
\cmidrule(lr){2-3}\cmidrule(lr){4-5}\cmidrule(lr){6-10}\cmidrule(lr){11-12}
System & Cert. & Heur. & Cert. & Heur. & TP & FP & TN & FN & Unk. (FP, FN) & all & clean & CPU (h) \\
\midrule
\code{VEB-Check} & 0 & 3{,}195 & 453 & 0 & 3{,}087 & 108 & 453 & 0 & 26, 236 & 0.6569 & 0.8837 & 2.1 \\
\code{ProofPulse} & 0 & 3{,}040 & 0 & 805 & 2{,}873 & 167 & 412 & 393 & 8, 57 & 0.4881 & 0.5228 & 4.2 \\
\midrule
\tool{} & 3{,}055 & 292 & 515 & 32 & 3{,}286 & 61 & 526 & 21 & 0, 16 & 0.9004 & 0.9226 & 58.0 \\
\bottomrule
\end{tabular*}
\end{table*}

\subsubsection*{Automatic certification (E5).} \tool{} automatically produces certificates for \numCertAutoNoSample{} of the \numCorpusTotal{}  methods in our dataset (\numCertAutoNoSamplePct{}\%). Of the \numCorpusStrong{}  specifications ultimately certified strong, \numStrongAuto{} (\numStrongAutoPct{}\%) are certified automatically by \tool{}. Of the \numWeakAutoOf{} specifications ultimately certified weak, \numWeakAuto{} (\numWeakAutoPct{}\%) are certified automatically by \tool{}. The remaining \numUncertified{} methods require additional effort to construct a certificate (E6), as discussed in \Cref{sec:discussion}.

\subsubsection*{Comparison with existing approaches (E4).}
We compare \tool{} with \veqcheck{} ~\cite{VeriEquivBench} and ProofPulse~\cite{ProofPulse_paper}, two existing approaches for assessing specification strength. \veqcheck{} attempts to verify that a specification is strong by performing an equivalence check against the reference implementation.
A successful check provides evidence that the specification is strong, whereas a failed check does not imply that the specification is weak. ProofPulse instead classifies specifications using a heuristic approach based on proof coverage and does not produce a certificate of strength or weakness under our definition. In contrast, \tool{} produces independently checkable certificates for both weak and strong specifications.

Table~\ref{tab:rq1matrix} summarizes the classifications produced by the three approaches, distinguishing those backed by a certificate (\emph{Cert.}) from heuristic classifications (\emph{Heur.}) and cases for which no classification is returned (\emph{Unk.}, divided into those that count as false positives and false negatives).  A false positive (FP) is a specification classified as weak but certified strong, whereas a false negative (FN) is classified as strong but certified weak. Methods without a classification are counted as FNs if their certified label is weak and as FPs if their certified label is strong. Both are reported in the \emph{Unk.} column. \veqcheck{} provides evidence-backed strong classifications for \numBaselineSureStrong{} methods, while its remaining classifications are heuristic, i.e., if a specification is not proven to be strong, it is considered weak by default.
ProofPulse provides only heuristic classifications. In contrast, \tool{} produces certificates for \numCertAutoNoSample{} methods (\numCertAutoNoSamplePct{}\%). For methods where no certificate is found, \tool{} falls back on the mutation score, using a threshold $\tau=1$ to classify a specification as strong when its score is at least $\tau$ and weak otherwise. In other words, the presence of any alive mutant causes the specification to be classified as weak. This is the most conservative and easily defensible threshold, since if the mutation tool could guarantee that it never produces equivalent mutants, finding another program that verifies would be sufficient to prove that the specification is weak. 
This produces heuristic classifications for \numReducedHeur{} additional methods. \tool{} returns no classification for the remaining \numReducedError{} methods (the \emph{Unk.} column of \Cref{tab:rq1matrix}), all of which are ultimately certified weak. For \numResidueWeakGeneratorError{} of them, the input generator fails before any mutant is run, because the Hypothesis strategy built from the preconditions is unsatisfiable or impractical, e.g., a recognized lower bound exceeds the default value range, or a length constraint demands a list with tens of thousands of elements. For the others, MutDafny generates no mutant to verify, so there is neither a certificate nor a mutation score to classify by. MCC(all) counts all of them as misclassifications. We analyze the accuracy of these classifications against the certified labels below.

Producing certificates requires additional computation, shown in the last column of \Cref{tab:rq1matrix} (\Cref{sec:eval:setup}). \tool{} uses \numReducedCpuHours{} hours of summed per-method CPU time, compared with \numBaselineCpuHours{} hours for \veqcheck{} and \numProofPulseCpuHours{} hours for ProofPulse.
Most of this time is spent on the methods that end up weak. The \numFigTimeStrongMethods{} methods proved strong take \numFigTimeStrongHours{} hours in total (median \numFigTimeStrongMedianSeconds{}\,s per method), and the \numFigTimeWeakMethods{} certified weak take \numFigTimeWeakHours{} hours (median \numFigTimeWeakMedianSeconds{}\,s), most of it spent verifying mutants. 
The \numFigTimeUncertifiedMethods{} methods left uncertified account for \numFigTimeUncertifiedHours{} hours, or \numFigTimeUncertifiedSharePct{}\% of the total execution time. Of these, 97 exhaust the per-method search budget.

\subsubsection*{Classification accuracy (E4--E6).}
We use the completed certified dataset as ground truth to evaluate the classifications produced by \tool{}, \veqcheck{}, and ProofPulse. The right-hand side of~\Cref{tab:rq1matrix} reports the resulting confusion matrices and Matthews correlation coefficient (MCC), the correlation between predicted and certified labels, which ranges from -1 (all wrong) through 0
(chance) to 1 (all correct) and, unlike accuracy, is not inflated by the dominance of weak specifications in our dataset.  We report MCC over the complete dataset, where unanswered cases are treated as incorrect, and over the subset of \numCleanSubset{}  methods for which all three approaches return a classification (\emph{MCC(all)} and \emph{MCC(clean)} in \Cref{tab:rq1matrix}).

\tool{} achieves an MCC of \numReducedMccAll{} over the complete dataset, substantially higher than \veqcheck{} (\numBaselineMccAll{}) and ProofPulse (\numProofPulseMccAll{}). When considering
only methods for which all approaches return a classification, \tool{} reaches an MCC of \numReducedMccClean{}, compared with \numBaselineMccClean{} for \veqcheck{} and \numProofPulseMccClean{} for
ProofPulse. Thus, the additional evidence produced by \tool{} also results in more accurate classifications.

\begin{figure}[t]
\centering
\includegraphics[width=0.8\linewidth]{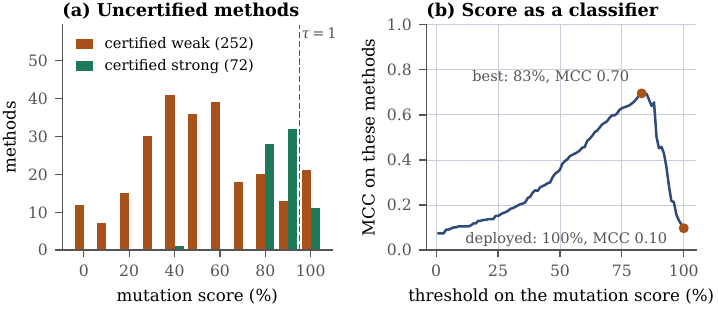}
\caption{The mutation score as a fallback classifier, on the \numScoreScored{} methods that \tool{} scores but does not certify. (a) Score distribution by certified label; methods left of the dashed line are classified weak under $\tau=1$. (b) MCC of the rule ``weak iff score $<$ threshold'' on these methods, for every threshold.}
\label{fig:score}
\end{figure}

\Cref{fig:score} shows what the mutation-score fallback contributes. On the \numScoreScored{} methods that \tool{} scores but does not certify, the scores of strong and weak specifications overlap. Equivalent mutants keep \numFigScoreStrongBelow{} of the \numScoreStrong{} strong specifications below a score of 100\%, so $\tau=1$ classifies almost all of them as weak, and \numFigScoreWeakFull{} weak specifications reach 100\% because no operator produces a mutant that verifies. On these methods, the score therefore reaches an MCC of only \numScoreThresholdMcc{} at $\tau=1$. A threshold of \numScoreBestThreshold{}\% would raise it to \numScoreBestMcc{} (AUC \numScoreAuc{}), but that threshold is tuned on the certified labels themselves. We keep $\tau=1$, which needs no tuning, and report heuristic classifications separately from certified ones.

\rqbox{\tool{} automatically certifies \numCertAutoNoSamplePct{}\% of the dataset. Automatic certification is more effective for weak specifications, reaching \numWeakAutoPct{}\%, than for strong specifications, reaching \numStrongAutoPct{}\%. Scored against the completed ground truth, its classifications reach an MCC of \numReducedMccAll{}, compared with \numBaselineMccAll{} for \veqcheck{} and \numProofPulseMccAll{} for ProofPulse.}

\subsection{RQ2: How do the certified labels produced by DafnyCert compare with those provided by existing benchmarks?}
\label{sec:rq2}

\subsubsection*{Disagreements with benchmark labels (E7).}
The certified labels disagree with the labels provided by the source benchmarks for \numCorpusChanged{} of the \numCorpusTotal{} methods in our dataset (\numCorpusChangedPct{}\%). As shown in~\Cref{tab:flips}, this includes \numVeribenchChanged{} of the \numVeribenchTotal{} VeriEquivBench methods (\numVeribenchChangedPct{}\%) and \numSynthChanged{} of the \numSynthTotal{} Dafny-Synthesis methods (\numSynthChangedPct{}\%).

Interpreting the VeriEquivBench disagreements requires some care because its labels are assigned at the file level, whereas our certificates classify individual methods. Of the \numVeribenchTotal{} VeriEquivBench methods, \numVbMultiMethods{} belong to files containing multiple methods and therefore inherit the same benchmark label. In particular, when a file is labeled weak, every specification in that file is marked as weak, even though some of those specifications may be strong under our definition. To avoid this mismatch, we restrict this analysis to the \numVbSingleFiles{} VeriEquivBench files containing a single method. Among these files, \numVbSingleToWeak{} of the \numVbSingleStrongTotal{} specifications labeled strong (\numVbSingleStrongChangedPct{}\%) are certified weak, whereas \numVbSingleToStrong{} of the \numVbSingleWeakTotal{} specifications labeled weak (\numVbSingleWeakChangedPct{}\%) are certified strong.

In the case of the Dafny-Synthesis benchmark, labels are human judgments over each method. Of the \numSynthChanged{} disagreements, \numSynthToWeak{} correspond to specifications labeled strong that we certify weak, while only one corresponds to a specification labeled weak that we certify strong.
The two disagreement directions have different interpretations, as discussed in~\Cref{sec:dataset}. A \emph{strong$\rightarrow$weak} disagreement provides definitive evidence that the specification admits multiple observable behaviors despite being labeled strong by the benchmark. Moreover, the weakness certificate includes a concrete alternative implementation and input witnessing this ambiguity. For tasks whose natural-language description determines a unique intended behavior, such a witness is sufficient to establish that the specification is weaker than intended and that the benchmark's strong label is incorrect. A \emph{weak$\rightarrow$strong} disagreement is less direct. The specification admits a unique behavior, but that behavior may still differ from the one described by the natural-language task. Resolving such cases therefore requires manual comparison with the task description.

Dafny-Synthesis contains one \emph{weak$\rightarrow$strong} disagreement. Manual inspection confirms that the specification determines a unique behavior, as indicated by our certificate, but that this behavior does not match the natural-language description. Thus, in this case, the original \emph{weak} label remains appropriate with respect to the intended task, even though the specification is strong in the behavioral sense measured by \tool{}.

\subsubsection*{Review of sampled VeriEquivBench disagreements (E10).}
VeriEquivBench has too many \emph{weak$\rightarrow$strong} disagreements to inspect them all, so we inspect a random sample of \numWeakToStrongReviewed{} of its \numVbSingleToStrong{} single-method cases. Two authors independently compared each specification with the natural-language description of its task. They agreed on all \numWeakToStrongAgreed{} methods, and in all \numWeakToStrongLabelWrong{} the specification matches the description. In every sampled case, the benchmark's \emph{weak} label is therefore wrong, and the certified \emph{strong} label is correct.

The way VeriEquivBench builds its labels explains this outcome. For each task, an LLM first translates the description directly into a \emph{strong} specification, which is not required to verify. In a second step, the LLM receives that specification, the description, and a reference Python solution, and writes verified Dafny code whose specification is labeled \emph{weak}. The label therefore records which step produced a specification, not an assessment of the specification itself. In \numWeakToStrongSameAsStrongSpec{} of the \numWeakToStrongReviewed{} sampled cases, the \emph{weak} specification has the same postconditions as the task's \emph{strong} one. In \numWeakToStrongStrongSpecNoEnsures{} of the others, the \emph{strong} specification has no postcondition at all, so the specification labeled \emph{weak} is the stronger of the two. The sampled methods, their certificates, task descriptions, and both reviewers' answers are part of the replication package.

\rqbox{Our certified labels reveal substantial errors among specifications labeled strong in both LLM- and human-labeled benchmarks, highlighting the difficulty of assessing specification strength. Among single-method VeriEquivBench files, whose labels are LLM-generated, \numVbSingleStrongChangedPct{}\% of specifications labeled strong are certified weak. In Dafny-Synthesis, whose labels are assigned by human annotators, \numSynthToWeak{} specifications labeled as strong are also certified as being weak. In the other direction, the benchmark's \emph{weak} label was wrong for all \numWeakToStrongReviewed{} sampled VeriEquivBench specifications that we certify strong.}

\subsection{RQ3: How do the strategies used to generate weakness certificates affect certification effectiveness and computational cost?}
\label{sec:rq3}

\tool{} uses two strategies to reduce the cost of finding certificates to show that a specification is weak, namely, guided input generation from method preconditions and a reduced set of mutation operators. We evaluate the contribution of each strategy and also measure the impact of the new nine mutation operators added to MutDafny.

\subsubsection*{Effect of precondition guidance (E8)}
To isolate the effect of precondition guidance, we ran the pipeline's input-generation step twice over the whole dataset. Once using the guided composite strategy of \Cref{tab:recognise}, and once using the plain per-type proposal. The guided proposal produces at least one input satisfying the precondition for
\numAblationGuided{} methods (\numAblationGuidedPct{}\%), compared with
\numAblationUnguided{} (\numAblationUnguidedPct{}\%) without guidance.
In particular, \numAblationGuidedOnly{} methods yield at least one admissible
input only with guidance.
The median fraction of accepted draws is
\numAblationGuidedAcceptPct{}\% with guidance, compared with
\numAblationUnguidedAcceptPct{}\% without it, since the guided proposal usually
satisfies the precondition by construction. Moreover,
\numAblationGuidedFull{} methods reach the full
\numAblationMaxExamples{}-input target with guidance, compared with
\numAblationUnguidedFull{} without it. Consequently, without guidance,
\numAblationGuidedOnly{} methods would receive no witness search because no
admissible input would be found for them.

\begin{table}[t]
\caption{The cheapest operator set reaching each target coverage $p$ on \textsf{Selection Set} (upper). \emph{Obtained $p$} is the coverage of \textsf{Selection Set}'s witness-bearing methods the set reaches, which can exceed the target. \emph{Certs.} is the \% of the ground truth certificates covered. \emph{Budget} is the share of the 112,734 mutants the full catalogue generates on \textsf{Selection Set}. $^{\dagger}$~is the target the dataset sweep ran at. \emph{Operators} names the operators we added and counts the rest as MO, operators of the original MutDafny catalogue.
}
\label{tab:operators}
\footnotesize
\setlength{\tabcolsep}{3pt}
\begin{tabular*}{\columnwidth}{@{\extracolsep{\fill}}lrrrrl@{}}
\toprule
Target $p$ & Obtained $p$ & Certs. \% & Budget \% & Ops & Operators \\
\midrule
\multicolumn{6}{@{}l}{\emph{Selection} --- cheapest set reaching each target $p$ on \textsf{Selection Set} (404 witness-bearing of 431 certified weak)} \\
1.000$^{\dagger}$ & 1.000 & 93.7 & 22.7 & 12 & {\tiny\ttfamily CAS, CES, INC, VSR} {\tiny + 8 MO} \\
0.990 & 0.990 & 92.8 & 11.8 & 9 & {\tiny\ttfamily CAS, CCU, INC, VSR} {\tiny + 5 MO} \\
0.950 & 0.950 & 89.1 & 4.5 & 8 & {\tiny\ttfamily CAS, CES, VSR} {\tiny + 5 MO} \\
0.900 & 0.901 & 84.5 & 1.4 & 2 & {\tiny\ttfamily VSR} {\tiny + 1 MO} \\
0.800 & 0.822 & 77.0 & 1.0 & 2 & {\tiny\ttfamily CES, VSR} \\
0.500 & 0.733 & 68.7 & 0.6 & 1 & {\tiny\ttfamily VSR} \\
\midrule
\multicolumn{6}{@{}l}{\emph{Added operators} --- cheapest set at 97.5\% coverage, the highest the original operators reach} \\
with added & 0.975 & 91.4 & 8.0 & 7 & {\tiny\ttfamily CAS, CCU, VSR} {\tiny + 4 MO} \\
original only & 0.975 & 91.4 & 52.5 & 8 & {\tiny 8 MO} \\
\midrule
\multicolumn{6}{@{}l}{\emph{Transfer} --- the chosen set, scored against each dataset's weak certificates} \\
\textsf{Transfer Set} & --- & 91.6 & --- & 12 & {\tiny of 2{,}841 weak, not fitted on at all} \\
\datasetSmall{} & --- & 98.0 & --- & 12 & {\tiny of 51 weak, operators' design dataset} \\
\bottomrule
\end{tabular*}
\end{table}

\subsubsection*{Mutation operator reduction (E1, E2)}
Evaluating every mutation operator is expensive because each generated mutant must be verified by Dafny. We therefore select a smaller operator set using the minimum-cost coverage formulation of~\Cref{sec:restricting_mutant_set}.
We perform operator selection on the \textsf{Selection Set} and measure how the selected set transfers on the \textsf{Transfer Set} (\Cref{sec:dataset}).

Of the \numSampleWeak{} specifications in the \textsf{Selection Set} that are ultimately certified weak, a version of \tool{} that uses all operators without early stop finds witnesses for \numChosenFiles{}. This run also has three times the deployed per-method budget, because it records every witness each mutant finds rather than stopping at the first. It is the record the operator selection is solved on, while the comparison of the two operator sets under the deployed budget and early stop is reported below (E3).
Table~\ref{tab:operators} reports the minimum-cost operator sets obtained for different coverage targets. Requiring complete coverage of the \numChosenFiles{}
methods selects \numChosenOps{} operators. This set retains \numChosenCoverage{}\% of the witnesses found by the version that uses all operators while generating only \numChosenPct{}\% of its mutants.

\subsubsection*{Effect of the operator reduction (E3).}
To isolate the effect of the operator reduction, we ran two experiments that
differ only in the operator set, using the same \numAbFiles{} methods from the
\emph{Selection Set} and the configuration in which the pipeline is
deployed, once with every operator and once with the selected set.
The reduced set requires \numAbReducedHours{}\,h of summed per-method time,
compared with \numAbUnrestrictedHours{}\,h for the unrestricted set, a
\numAbSpeedup{}$\times$ speedup. The mutation stage alone requires
\numAbReducedMutationHours{}\,h, compared with
\numAbUnrestrictedMutationHours{}\,h.
Unexpectedly, the reduced experiment also finds \emph{more} witnesses,
\numAbReducedWitnesses{} compared with \numAbUnrestrictedWitnesses{}.
Since removing operators cannot increase the set of methods they can cover,
this difference is due to the fixed per-method budget, since
\numAbUnrestrictedTimeout{} methods exhaust their budget with the unrestricted
set, compared with \numAbReducedTimeout{} with the reduced set. Per method, both experiments find witnesses for \numAbBothWitnesses{} methods. The reduced set finds \numAbGained{} more, every one of them a method on which the unrestricted set exhausted its budget, and misses \numAbLost{} that the unrestricted set finds, \numAbLostToTimeout{} of them because it exhausts its own budget.

\subsubsection*{Transfer to the \textsf{Transfer Set} (E9)}
\Cref{tab:operators}'s \emph{Certs.} column gives, for each row, the fraction of a dataset's weak certificates that the operator set recovers with a witness.
The denominator is the set of methods that the dataset ultimately certifies as weak. 
The selected operators cover \numChosenCertsPct{}\% of weak specifications in the \textsf{Selection Set} and find witnesses for \numGtVbRecall{}\% in the \textsf{Transfer Set} under the deployed configuration. This suggests that the selected operators generalize well to methods not used during selection.
The \textsf{Transfer Set} is independent of the selection, because
the selection saw only the \textsf{Selection Set}, and the nine
operators added to MutDafny's catalogue were designed by inspecting
\datasetSmall{} (\Cref{sec:arch:weak}).
\datasetSmall{} reaches \numGtSynthRecall{}\%, but since the added operators were designed on this dataset, this figure does not show generalization.

\subsubsection*{Contribution of the added operators (E1, E2).}
Four of our nine additions survive selection (\Cref{tab:operators}) and make up a third of the operators used in the dataset sweep. They contribute in two ways: they reach methods that no operator of the original MutDafny catalogue reaches, and they reach the remaining methods with far fewer mutants.

\emph{Coverage.} We measure generalization on the \textsf{Selection Set}, since no added operator was designed on it. There, the original MutDafny catalogue alone finds a witness for \numPublishedSampleCovered{} of the \numChosenFiles{} witness-bearing methods (\numPublishedSamplePct{}\%), so the additions are necessary for only \numAddedOnlySample{} (\numAddedOnlySamplePct{}\%). On \datasetSmall{}, on which they were designed, they are necessary for \numAddedOnlyHeldOut{} of the \numHeldOutFiles{} weak \datasetSmall{} methods (\numAddedOnlyHeldOutPct{}\%), almost all through \code{CAS} and \code{CES}. Judged by coverage alone, the additions therefore have a modest impact.

\emph{Cost.} The main benefit of the added operators is a lower cost. Without them, no set of operators covers more than \numPublishedFloor{}\% of the witness-bearing methods in the \textsf{Selection Set}, so the deployed selection at $p=1$ would have no solution. We therefore compare both catalogues at \numPublishedFloor{}\% coverage, the highest coverage they both reach (middle block of \Cref{tab:operators}). The cheapest set of original operators needs \numPublishedCoverOps{} operators and \numPublishedCoverPct{}\% of all mutants. With the added operators, \numMatchedCoverOps{} operators and \numMatchedCoverPct{}\% of the mutants are enough to reach \numPublishedFloor{}\%  coverage. The same coverage thus costs \numPublishedCostRatio{}$\times$ fewer mutants. The original catalogue has no cheap operator that changes the contents of a collection. Its cheapest cover must therefore use broad operators such as \code{EVR}, \code{ROR}, and \code{MRR}, which mutate almost every method and so produce many mutants that must all be verified. Because verifying mutants dominates the running time (\Cref{sec:rq1}), this reduction matters more than the small gain in coverage.

\rqbox{Guided input generation increases the fraction of methods for which admissible inputs are found from \numAblationUnguidedPct{}\% to \numAblationGuidedPct{}\%. 
The selected \numChosenOps{} operators generate only \numChosenPct{}\% of the mutants on the \textsf{Selection Set} and find witnesses for \numGtVbRecall{}\% of weak specifications on the \textsf{Transfer Set}. They also reduce execution time by $\numAbSpeedup{}\times$ on the \textsf{Selection Set}. The added operators make the $p=1$ cover feasible and cut the mutants needed at \numPublishedFloor{}\% coverage by \numPublishedCostRatio{}$\times$.
}

\subsection{Threats to Validity}
\label{sec:threats}

\subsubsection*{Trusted base} Certified labels rely on Dafny (verifier, Z3, and, for weak certificates, the compiler and its runtime precondition checks) and the validator, which checks (V1)--(V3) and (V6) itself. Certificate producers (\tool{}, humans, and the LLM agent) are not trusted. We assume that $M$ is deterministic. Strength certificates assume that the precondition is satisfiable. The validator checks for vacuity (V5) and rejects certificates when Dafny detects it, but this check cannot rule out all vacuous cases.

\subsubsection*{External validity} The dataset is the largest publicly available
collection of labelled Dafny specifications, but it is not industrial code, since
both sources are benchmark problems with self-contained methods and small
signatures. Industrial Dafny has larger heap footprints and richer module
structure, which we expect to make applying our approach harder in engineering terms.

\subsubsection*{Unsupported constructs} Self-referential reference types are out of
scope by construction and are
filtered at dataset-creation time (\Cref{sec:dataset}). In addition, the witness search's precondition pattern table is
finite, and existentials over domains the compiler cannot enumerate are handled
only in the common case. A precondition shape we do not model can leave a method uncertified, but it cannot produce a wrong certificate, as the validator step checks this.

\section{Ground Truth Completion and Failure Analysis}
\label{sec:discussion}

\tool{} automatically produces certificates for \numCertAutoNoSample{} of the \numCorpusTotal{} methods in our dataset (\numCertAutoNoSamplePct{}\%), leaving \numUncertified{} methods uncertified. These uncertified methods can, in principle, be certified manually. For a weak specification, one can provide an alternative implementation that still verifies against the original contract, together with an input on which the two implementations disagree. For strong specifications, one can give additional information to the harness such as auxiliary assertions, loop annotations, or lemmas that will help Dafny to verify that no alternative output is possible.
In this section, we describe how we complete the ground truth for the remaining methods and then use the resulting certificates to analyze why \tool{} failed to certify them.

\subsection{Completing the Ground Truth}
\label{sec:disc:groundtruth}

Manually constructing certificates for all \numUncertified{} uncertified methods would require substantial effort. We therefore first solved a small number of cases manually to identify recurring strategies for constructing weakness witnesses and completing strength proofs. We then encoded these strategies in a detailed prompt (available in the replication package), together with \numAgentPromptExamplesTotal{} manually completed certificates as few-shot examples (\numAgentPromptExamples{} dataset methods and three small illustrative programs that are not in the dataset), and used a Claude Code agent (version 2.1.277) running Claude Sonnet 5 to apply them to the remaining \numAgentClosed{} methods. 

Since the solutions of the \numAgentPromptExamples{} dataset examples appear in the prompt, we closed those methods by hand and did not give them to the agent, which therefore closed the other \numAgentClosed{} of the \numUncertified{} uncertified methods.
The LLM agent also had access to a script that generates the skeleton of a strength or weakness certificate, which it then completed with either the required proof information or a candidate alternative implementation and witness.

The LLM agent is not part of the trusted base because every candidate certificate must pass the same validator used for \tool{}'s certificates (\Cref{sec:mot:validator}).  
Therefore, the correctness of the resulting label depends on the certificate passing validation, rather than on whether it was produced automatically, manually, or by an LLM agent.

The validator accepted certificates for all \numAgentClosed{} methods processed by the LLM agent. The agent could invoke the validator itself and repair rejected candidates. Of the \numAgentClosed{} certificates, \numAgentFirstSubmissionPass{} passed on the first validator attempt, and no method required more than \numAgentMaxValidatorCalls{} attempts. Per method, the agent performed a median of \numAgentMedianTurns{} actions, including reading the method, generating the certificate template, editing it, and invoking the validator. Together with the four methods closed by hand, this yields certificates for all \numCorpusTotal{} methods in the dataset.

\tool{} needs \numToolCpuHours{} hours of summed per-method time and produced certificates for \numCertAutoNoSamplePct{}\% of the methods. Each method is certified independently and the only resource used is CPU, so this time divides almost evenly across processes, which is about \numToolCpuHoursPerProcess{} hours on the \numToolConcurrency{} processes of our machine. The LLM agent completed the remaining \numAgentClosed{} methods in around \numAgentWallHours{} hours, a time that does not shrink as easily with more processes, because the agent is bound by the API's rate limits and per-token cost rather than by local CPU, at a total cost of \$\numAgentCostUsd{}, or \$\numAgentCostPerMethodUsd{} per method. Restricting the LLM agent to methods that remain uncertified substantially reduces the cost of obtaining a complete ground truth. Extrapolating from the observed cost on the uncertified methods, applying the same LLM agent directly to all \numCorpusTotal{} methods would cost approximately \$\numAgentFullCorpusUsd{} and take about \numAgentFullCorpusHours{} hours. In contrast, \tool{} incurs no per-method LLM cost, and certifies the large majority of methods in substantially less time.

\subsection{Weak Specifications Not Certified Automatically}
\label{sec:disc:weak}

\begin{table}[t]
\centering
\footnotesize
\caption{Reasons why \tool{} fails to automatically certify the remaining \numResidueWeak{} weak specifications.}
\label{tab:weak-failures}
\begin{tabular}{lr}
\toprule
Reason & Methods \\
\midrule
Input generator failed
    & \numResidueWeakGeneratorError{} \\
No mutants generated
    & \numResidueWeakNoMutants{} \\
No admissible inputs found
    & \numResidueWeakNoInput{} \\
No divergence ($<50$ admissible inputs)
    & \numResidueWeakStarved{} \\
No divergence ($\geq 50$ admissible inputs)
    & \numResidueWeakAmpleButSilent{} \\
\midrule
Total & \numResidueWeak{} \\
\bottomrule
\end{tabular}
\end{table}

Of the methods left uncertified by \tool{}, \numResidueWeak{} are ultimately certified weak. Table~\ref{tab:weak-failures} summarizes the reasons why these methods were not certified automatically.

For \numResidueWeakStarvedNoSearch{} methods, \tool{} produces no heuristic classification because input generation fails for \numResidueWeakGeneratorError{} methods and MutDafny generates no mutants for the remaining \numResidueWeakNoMutants{} (\cref{sec:rq1}).
For \numResidueWeakNoInput{} methods, input generation succeeds but finds no admissible input within the available budget. A likely cause is a precondition shape not covered by the patterns in Table~2, such as an existential or a recursive predicate. In these cases, the generator falls back to filtering, which is unlikely to satisfy restrictive preconditions.

For another \numResidueWeakNoDivergence{} methods, the search finds admissible inputs but no divergence. We use 50 inputs as a threshold to distinguish limited input generation from more extensive sampling. Of these, \numResidueWeakStarved{} produce fewer than 50 admissible inputs. The remaining \numResidueWeakAmpleButSilent{} produce at least 50 admissible inputs and alive mutants, but none of the sampled inputs exposes a divergence. This may indicate insufficient input coverage or mutants that are behaviorally equivalent to the original implementation.

Overall, many of the remaining failures come from the difficulty of finding inputs that expose a behavioral difference between the original implementation and a mutant. This suggests moving from largely random sampling toward more targeted witness generation. Inputs could, for example, be directed toward branch conditions affected by a mutation or boundary values appearing in the specification. Similarly, mutants could be constructed using values admitted by the specification rather than relying only on fixed syntactic transformations.

\subsection{Strong Specifications Not Certified Automatically}
\label{sec:disc:strong}

The remaining \numResidueStrong{} methods are certified strong. For all of them, Dafny fails to verify the equivalence harness generated by \tool{}. These failures occur quickly, with a median verification time of \numResidueStrongProofSeconds{} seconds, suggesting that increasing the verification budget would not resolve them. Instead, the harnesses require additional proof information that Dafny cannot infer automatically.

The completed certificates show a consistent pattern. The added proof information has a median size of \numResidueStrongProofLines{} lines and at most \numResidueStrongProofLinesMax{} lines. Of the \numResidueStrong{} certificates, \numResidueStrongLemma{} introduce a lemma, and every such lemma is recursive. In addition, \numResidueStrongAssert{} use intermediate assertions and \numResidueStrongLoopAnnotation{} use additional loop annotations. These categories are not mutually exclusive. The dominant difficulty therefore appears to be proving properties that require inductive reasoning, for example over sequences, array indices, or recursive datatypes.

These results suggest that improving the certification of strong specifications requires better proof synthesis rather than a larger verification timeout. 
In particular, automatically generating recursive lemmas, intermediate assertions, and loop annotations could allow \tool{} to certify many of these specifications without additional proof construction.

\rqbox{
The failure analysis exposes different bottlenecks. Weak specifications are missed primarily because the search fails to find a suitable witness, whereas strong specifications are missed because their equivalence proofs require additional inductive reasoning.
}

\section{Related Work}
\label{sec:related}

\subsubsection*{Judging specification strength.} 

Clover \cite{Clover} combines LLMs and formal verification to check consistency between code, specifications, and natural-language descriptions.
Mirchev et al. \cite{mirchev} jointly refine code, specifications, and tests to improve their consistency.
IronSpec \cite{ironspec} uses sanity checks, specification mutations, and proofs to identify specification bugs.
In contrast, \tool{} produces checkable certificates showing whether a specification uniquely determines program behavior.

\citet{endres2024} and \citet{lahiri2024} propose metrics for measuring the strength of LLM-generated specifications. \citet{endres2024} quantify completeness as the fraction of mutants that fail under the specification, using likely faulty LLM-generated implementations and filtering out equivalent mutants with a test suite. Any mutant that fails a test is guaranteed to be non-equivalent. \citet{lahiri2024} simplifies this by mutating the outputs of the validation test suite. Verus-SpecGym~\cite{agarwal2026verus} extends this framework to an agentic setting, evaluating generated Verus specifications against human-written test cases. These approaches rely on a test suite, which, in practice, is often missing in verification-aware contexts, and there are no guarantees that the test suites themselves are complete. 

MutDafny~\cite{Mutdafny} generates mutants deterministically, flagging alive ones as hints of a potential specification weakness, without needing a test suite. However, it produces many equivalent mutants, requiring a lengthy manual analysis of the alive mutants to confirm the weaknesses. \citet{le-cong2025} also use mutations to measure the completeness of Java Modeling Language (JML) specifications, but their completeness rate remains a heuristic rather than a definitive verdict.

ProofPulse \cite{ProofPulse_paper} inspects a correctness proof and classifies a specification as strong if every line of the implementation is used to prove the postcondition. 
However, this can be misleading, as a specification may admit divergent implementations even when every line of the proof is used. 

\veqcheck{} ~\cite{VeriEquivBench}, the equivalence check of ~\datasetBig{}, determines if a specification is strong through an equivalence proof harness. Successful verification provides assurance that the specification is strong, but a failure does not necessarily indicate a weakness, as strong specifications may still fail the check. This approach inspired our strength check, which we improved to reduce the cases of non-verification for strong specifications. To distinguish genuinely weak specifications from strong ones for which the verifier cannot complete the proof, we complement the strength check with a weakness certification approach. This combination allows us to produce certificates in both directions, which can be re-checked independently using Dafny and our validator, without trusting the generation pipeline.

\subsubsection*{Test generation for Dafny.} Before building our Hypothesis frontend, we considered existing test generation options for Dafny. Dafny itself provides built-in test generation through DTest \cite{DTest}, which does not support array inputs and preconditions with nested quantifiers. Moreover, this tool only generates the minimum number of tests required to achieve program coverage, which may be insufficient to catch divergent implementations. DSpec2Test \cite{DSpec2Test} improves this to allow the generation of multiple test cases per program path, but its test generation approach is guided by the specification, rather than by the implementation, and it still suffers from lack of support for some Dafny features. DafnyCBT \cite{DafnyCBT} is another specification-based test generation tool, which relies on its own Dafny-to-SMT translation. 

We avoid solver-based generation primarily because of speed and reliability. Much like our sampling approach, the extracted test inputs have to satisfy preconditions, and the SMT calls needed to extract them are much slower than our constructive approach. 

Furthermore, Z3 is difficult to steer towards varied inputs and tends to return the simplest ones. The main drawback of both DSpec2Test and DafnyCBT is, however, that they generate tests from the specification alone. Therefore, they are not fit for our use-case, which is to test the specification itself. 

Our frontend follows the property-based testing tradition of QuickCheck~\cite{quickcheck} and Hypothesis~\cite{Hypothesis}, drawing inputs from generators built from the precondition.

\section{Conclusion}
\label{sec:conclusion}
A verified Dafny method is only as trustworthy as its specification. We presented \tool{}, which replaces heuristic assessments of specification strength with independently checkable certificates. A strength certificate uses a verified equivalence harness to establish that a specification admits at most one observable behavior for each admissible input, while a weakness certificate provides a verified alternative implementation and an input on which the two implementations disagree.

Applied to \datasetBig{} and \datasetSmall{}, \tool{} automatically certifies \numCertAutoNoSamplePct{}\% of the \numCorpusTotal{} methods. Guided input generation and a reduced set of \numChosenOps{} mutation operators make weakness certification more efficient. The selected operators generate only \numChosenPct{}\% of the mutants on the Selection Set while finding witnesses for \numGtVbRecall{}\% of weak specifications on the held-out Transfer Set. Certificates for the remaining \numUncertified{} methods also passed the same validator, providing recheckable certificates for all \numCorpusTotal{} methods.

The certified labels disagree with the benchmarks' original labels for \numCorpusChangedPct{}\% of the methods. Among single-method \datasetBig{} files, \numVbSingleStrongChangedPct{}\% of specifications labeled strong are certified weak, as are 19 specifications labeled strong in the human-annotated \datasetSmall{} benchmark.
These findings highlight the limitations of relying on human- or LLM-assigned strength labels without independently checkable evidence, particularly when such labels are used to evaluate specification generation techniques.

The failure analysis identifies two main directions for improvement. 
Weakness certification is primarily limited by the difficulty of finding admissible inputs and behavioral divergence, motivating more targeted witness generation.
Strength certification often requires additional inductive reasoning, motivating automated synthesis of lemmas and invariants. The certificate-based approach could also be extended to other verification-aware languages, providing independently checkable evidence of specification strength beyond Dafny.

\section*{Data-Availability Statement}

\tool{} and its replication package are open source under the MIT license and archived on Zenodo at \url{https://zenodo.org/records/23326759}.
The archive contains a Docker image with everything needed to run \tool{} and reproduce the results.
Besides the image, the archive contains a reviewer guide (\code{README.md}), the prompt of the agent fallback (\code{agent\_prompts/}), and the supplementary material (\code{supplementary\_material/}).

\begin{acks}

Alexandra Mendes and Isabel Amaral were partially funded by National Funds through the FCT - Fundação para a Ciência e a Tecnologia, I.P. (Portuguese Foundation for Science
and Technology) within the project VeriFixer, with reference 2023.15557.PEX (DOI: 10.54499/2023.15557.PEX) and by an Amazon Research Award, Fall 2024. 

Ruben Martins was partially supported by the National Science
Foundation (NSF) under Award CCF2427581 and DARPA Agreement FA8750-24-9-1000.

Álvaro Silva was co-financed by National Funds through the FCT - Fundação para a Ciência e a Tecnologia, I.P. (Portuguese Foundation for Science
and Technology) within the project VeriFixer, with reference 2023.15557.PEX (DOI: 10.54499/2023.15557.PEX), and Fundação para a Ciência e a Tecnologia (Portuguese Foundation for Science and Technology) through the Carnegie Mellon Portugal Program under the fellowship reference PRT/BD/155045/2024.

\end{acks}

\bibliographystyle{ACM-Reference-Format}
\bibliography{software}

\end{document}